\documentclass[11pt,a4paper]{article}
\pdfoutput=1
\usepackage{jheppub}	

\usepackage[normalem]{ulem}	

\newcommand{\blue}[1]{\color{blue} #1 \color{black}}

\usepackage{graphicx}
\usepackage{multirow}

\def\lsim{\raise0.3ex\hbox{$\;<$\kern-0.75em\raise-1.1ex
\hbox{$\sim\;$}}}
\def\gsim{\raise0.3ex\hbox{$\;>$\kern-0.75em\raise-1.1ex
\hbox{$\sim\;$}}}

\title{
\blue{Possible Interpretations of IceCube High-Energy Neutrino Events}
}
\author{Chee Sheng Fong$^{1}$}
\author{Hisakazu Minakata$^{1,2}$}
\author{Boris Panes$^{1}$}
\author{and Renata Zukanovich Funchal$^{1,2}$}

\affiliation{
$^1$Instituto de F\'{\i}sica, Universidade de S\~ao
  Paulo, C.\ P.\ 66.318, 05315-970 S\~ao Paulo, Brazil \\ 
$^2$Kavli Institute for Theoretical Physics, University of California
Santa Barbara, CA 93106-4030 USA
}
\emailAdd{fong@if.usp.br}
\emailAdd{minakata@fmail.if.usp.br}
\emailAdd{bapanes@if.usp.br} 
\emailAdd{zukanov@if.usp.br}

\abstract{ We discuss possible interpretations of the 37 high energy
  neutrino events observed by the IceCube experiment in the South
  Pole.  We examine the possibility to explain the observed neutrino
  spectrum exclusively by the decays of a heavy long-lived particle of
  mass in the PeV range. We compare this with the standard scenario,
  namely, a single power-law spectrum related to neutrinos produced by
  astrophysical sources and a viable hybrid situation where the
  spectrum is a product of two components: a power-law and the
  long-lived particle decays. We present a simple extension of the
  Standard Model that could account for the heavy particle decays that
  are needed in order to explain the data.  We show that the current
  data equally supports all above scenarios and try to evaluate the
  exposure needed in order to falsify them in the future.  }


\preprint{NSF-KITP-14-202}
\arxivnumber{arXiv:1411.5318}

\begin{document}
\maketitle

\section{Introduction}
\label{sec:intro}

The particle physics community around the globe invested a few decades
in a variety of experiments in a coordinated effort to scrutinize the
Standard Model (SM) of electroweak interactions. The scientific
achievements turned out to be impressive.  Neutrino flavour
oscillations have been discovered and confirmed by a string of solar,
atmospheric, reactor and accelerator neutrino experiments~\cite{osc},
challenging our understanding of neutrino properties and their role in
nature.  The discovery of the Higgs boson by the LHC experiments in
2012~\cite{LHC}, finally set the crowning glory of the SM
electroweak symmetry breaking mechanism to experimental probing.  The
IceCube neutrino observatory, located in the South Pole, reported in
2013 28 neutrino candidates in the energy range from 50 to 2000 TeV,
constituting at the time a 4.1$\sigma$ excess over the expected
atmospheric background~\cite{Aartsen:2013jdh}. This exciting
discovery of high-energy cosmic neutrinos heralds the dawn of neutrino
astronomy and drives intriguing questions.  Where do these neutrinos
come from?  How were they produced?  Can they shed a light on the long
lasting quest for the origin of cosmic rays?  Are they related to
the elusive Dark Matter (DM)?

Since this first announcement many authors have investigated the
possible origin of these events.  Among the possibilities explored are
viable astrophysical sources~\cite{Roulet:2012rv, Cholis:2012kq,
  Kalashev:2013vba, Stecker:2013fxa,
  Murase:2013rfa,Anchordoqui:2013qsi,Murase:2013ffa,
  Winter:2013cla,Ahlers:2013xia,Taylor:2014hya,Sahu:2014fua}, a
possible DM connection~\cite{Feldstein:2013kka,Bai:2013nga,Esmaili:2013gha,
  Bhattacharya:2014vwa,Bhattacharya:2014yha,Esmaili:2014rma}, 
a leptoquark resonance~\cite{Barger:2013pla},
the decay of massive neutrinos~\cite{Pakvasa:2012db,Higaki:2014dwa}, the decay of a
very heavy long-lived particle~\cite{Ema:2013nda,Ema:2014ufa} or even
the possibility that these events could be understood in terms of novel
interactions that neutrinos have with the cosmic neutrino
background~\cite{Ioka:2014kca,Ng:2014pca,Cherry:2014xra}.  Finally, in
Ref.~\cite{Chen:2013dza} the authors showed that assuming a simple
isotropic astrophysical power-law spectrum for the neutrinos, the
current data is consistent with neutrino having only the SM
interactions.\footnote{In Ref.~\cite{Chen:2014gxa}, the authors extended
the study to consider two-component astrophysical flux.}

Recently IceCube has updated their analysis to include three years of
data for a live time of 988 days. They have now 37 neutrino events
with energy from 30 TeV to 2 PeV~\cite{Aartsen:2014gkd}, strongly
disfavouring a purely atmospheric explanation at 5.7$\sigma$.
Interestingly enough, there are still no events in the energy range from
$\sim$ 400 TeV to 1 PeV. This may simply be a statistical fluctuation
or a signature from the underlying physics.  Furthermore, the energy
distribution indicates that there may be a cutoff in the spectrum at
PeV energies.

In this paper we discuss interpretations of the IceCube high-energy
excess events.  Among the candidates so far examined the following two
scenarios seem to be ``standard'' and relatively model-independent:
(a) a power-law spectrum to which astrophysical sources are kept in
mind, (b) decays of a heavy Long-Lived Particle (LLP), but with a
background power-law spectrum component.  We investigate here, in
addition to these two, the possibility that (c) decays of a heavy LLP
alone can explain the IceCube data in its entire energy region.  The
last possibility was also studied in
Refs.~\cite{Esmaili:2013gha,Bai:2013nga}.  
We emphasize here that the LLP we are considering can generically 
constitute only a fraction of the total DM.
With the current IceCube 3-year data \cite{Aartsen:2014gkd} and
using \emph{only} energy spectrum information, we will
show that all the three scenarios above (a) -- (c) can fit the current 
data well.  Although the analysis of the angular distribution of the events
showed a preference for a DM-like distribution over the
isotropic distribution, more data is required to draw a robust 
conclusion \cite{Bai:2013nga,Esmaili:2014rma}. Hence in this work,
we will only focus on energy spectrum information.
The next question, then, is how these three scenarios can be
distinguished.  We investigate this problem by examining simulated
data of IceCube for the future to establish the necessary exposure
time to differentiate the three scenarios above.

This paper is organized as follows. In Section~\ref{sec:LLPonly}, we
discuss the generic properties of the LLP that are required to explain
the features of the IceCube events. In Section~\ref{sec:anal}, we
describe the various sources of neutrinos in IceCube required for our
data analysis.  In Section~\ref{sec:fitting}, we perform an analysis
of the IceCube data in view of the three scenarios (a) -- (c).  In
Section~\ref{sec:future} we discuss how our model can be falsified
with future IceCube data. In Section~\ref{sec:model} we construct a
LLP model that could accommodate the IceCube observations.  We discuss
the LLP abundance in our model and conclude that it does not have to
be a dominant part of the DM content of the Universe. Finally, in
Section~\ref{sec:conc} we make our final remarks and conclusions. This paper 
is completed with two appendices: In Appendix~\ref{details}, 
we give additional details on the statistical treatment we employed in this work
while in Appendix~\ref{summary}, we list the confidence intervals of 
the best fit parameters for various scenarios considered in this study.

\section{Can a Decaying Long-Lived Particle Alone Explain the IceCube data?}
\label{sec:LLPonly}

The spectacular detection of 37 neutrino events in the energy range
from 30 TeV to 2 PeV within the three years data set of IceCube
disfavours, as already mentioned, a purely atmospheric explanation at
5.7$\sigma$~\cite{Aartsen:2014gkd}.  Moreover, neutrino events are
absent in the energy range from about 400 TeV to 1 PeV and beyond 2
PeV, while the assumption of an unbroken $E^{-2}$ flux,
would predict three events beyond 2 PeV~\cite{Aartsen:2014gkd}.  So
the non-observation of events at higher energies seem to suggest there is
a cutoff at the PeV scale. Although at the moment this is still
consistent with statistical fluctuations, here we opt for the exciting
possibility that these features could arise from the decays of a single 
LLP. Clearly, more than one species of decaying LLP with different
masses could as well describe these features.  We will take the minimal
point of view and stick to a single species of LLP in this work.

To accommodate the two features, the {\em dip} and the {\em cutoff},
described above, we need a decaying LLP, $Y$, with mass, $M_Y$, at the PeV
scale and lifetime, $\tau_Y$, with the following properties:
\begin{enumerate}
\item[(i)] \emph{It has to be long-lived.\footnote{This is probably
    obvious from its name.}} \\ In order for $Y$ to remain today, it
  has to have a lifetime at least longer than the lifetime of the
  Universe, {\em i.e}, $\tau_Y > t_0 \simeq 4.4 \times 10^{17}$
  s.\footnote{See, however, Refs.~\cite{Ema:2013nda,Ema:2014ufa} where
    they discussed a scenario where the decaying particle can have a
    lifetime shorter than the age of the Universe but its mass has to
    be very heavy, around $\sim 10^4$ PeV.}

\item[(ii)] \emph{It has a two-body decay to at least a SM neutrino in
  the final state.} \\ Assuming the LLP to be non-relativistic, the
  two-body decay $Y\to \nu_\alpha \, N$ with $\nu_\alpha$ a SM
  neutrino of flavour $\alpha = e,\mu,\tau$ and $N$ can be neutrino 
  or another SM singlet field, will produce a peak 
  in the energy spectrum at $M_Y/2$.  The
  cutoff in the energy spectrum will be at $M_Y/2$ as well.  As shown
  in Ref.~\cite{Esmaili:2013gha}, electroweak (EW) corrections can in
  fact soften the peak and give a low energy tail. 
  Intriguingly, they also showed that a peak
  structure can also result from the decay $Y \to \bar{e} \, e$ due EW
  corrections (e.g. cascade radiation of massive gauge
  bosons)~\cite{Kachelriess:2009zy,Ciafaloni:2010ti}.  In the current
  work, we take into account EW corrections in a simple way by
  extrapolating the results of \cite{Ciafaloni:2010ti} to higher
  masses.  We know this is not strictly correct as the EW corrections
  in that paper were computed at leading order and so their results
  are only valid as long as the particle is not too heavy. However, 
  the calculation of higher order corrections would be beyond the scope 
  of this work.

\item[(iii)] \emph{It has to admit at least another longer decay chain
  to neutrinos.}\\ In order to produce neutrinos with a continuum
  energy spectrum at lower energies, the LLP requires a longer decay
  chain to neutrinos, $Y \to ... \to \nu_\alpha \nu_\beta...$, as
  considered, for instance, in
  Refs.~\cite{Covi:2009xn,Bai:2013nga,Esmaili:2013gha}.  For example,
  if $Y$ is a scalar, we can have $Y \to \mu \bar\mu$, $Y \to \tau
  \bar\tau$, $Y \to t\bar t$, $Y \to 2h$ etc (we use the standard notation 
  where $h$ is the SM Higgs, $t$ for top and so on). 
  Using PYTHIA~\cite{Sjostrand:2007gs}, we generated neutrino energy spectra
  from the decay of the LLP at rest with its mass at the PeV
  scale. Experimenting with various channels, we found that in order
  to account for the excess of neutrino events at lower energies
  (where the atmospheric neutrinos alone cannot explain the data),
  two-body decays such as $Y \to 2h$ as well as four-body decays, 
  e.g. $Y\to t\bar t t \bar t$ or $Y \to 4h$, can
  do the job. The decay spectra will also be modified by EW
  corrections and this will be partially taken into account by us as
  described above (ii). In fact, these corrections, in general, tend
  to shift neutrino events to lower energies. As we will show in the
  work, even considering only (ii) $Y \to \nu_\alpha N$, we will also
  have a low energy tail due to EW corrections and red shift effect
  from extra-galactic contributions (see also \cite{Esmaili:2013gha}).

\end{enumerate} 

In Sections \ref{sec:anal} - \ref{sec:future}, we will consider the
LLP to be a scalar $Y$ with three possible decay channels: $Y \to
\nu_\alpha \,N$, $Y \to 2h $ and $Y \to 4h$. Here we assume $N$ is a 
fermion singlet which is not a SM neutrino. If the reader is
curious about its identity, he or she can go directly to Section
\ref{sec:model} where we present a consistent model which can realize
such a particle with the required decay channels.  For simplicity, we
will assume that $Y$ decays with equal branching ratios to neutrinos
of all flavours and suppress the flavour index in what follows, {\em
  i.e.} $Y \to \nu \, N$.  For the decays $Y \to 2h$ and $Y \to 4h$,
we use PYTHIA to generate the energy spectrum of the neutrinos taking
into account of SM particles decay and hadronization.\footnote{The EW
  corrections to the neutrino energy spectrum are taken into account
  only for the decay channels $Y \to \nu \, N$ and $Y \to 2h$ as
  described above (ii) while for the channel $Y \to 4h$, since there
  is no straightforward way to implement such corrections, we ignore
  them keeping in mind that these corrections tend to shift neutrino
  events to lower energies.}  Then we consider neutrino oscillations
to properly account for the neutrino flavours which arrive at the
Earth.  Finally before proceeding to data analysis, we want to stress
that the above choice is by no means the unique decay channels for a
LLP which can describe the suggestive features of the IceCube data but
simply a working assumption.

\section{IceCube Data Analysis: A Brief Description}
\label{sec:anal}

We will consider in our calculations three different sources 
of neutrinos in IceCube: LLP decays, an unknown astrophysical source 
(modelled with a power-law spectrum) and cosmic ray air showers.

The contributions of LLP decays to the IceCube neutrino flux have two 
different components: a galactic and a diffuse extra-galactic contribution. 
The neutrino differential flux from galactic LLP decays can be calculated 
as~\cite{Bai:2013nga} 

\begin{equation}
\frac{d\Phi_{\nu}}{dE_{\nu}\, db\, dl}=\frac{1}{N}\frac{dN}{dE_{\nu}}
\frac{1}{\tau_{Y}M_{Y}}\frac{\cos(b)}{4\pi}\int ds \, \rho_{Y}(r(s)),
\end{equation}
where the integral on $s$ is along the line of sight and
$r^{2}=s^{2}+r_{0}^{2}-2\, s\, r_{0}\cos(l)\cos(b)$, with
$-90^{\circ}\leq b<90^{\circ}$ and $-180^{\circ}\leq l<180^{\circ}.$
Here $r_{0}=8.5$ kpc is the distance from the Sun to the galactic
center. The term $(1/N)dN/dE_{\nu}$ is the normalized neutrino energy
spectrum calculated using PYTHIA in the context of the LLP scenario 
proposed. 
For the galactic LLP matter density we use the DM Einasto density profile,

\begin{equation}
\rho_{Y}(r)=\rho_{0}\, e^{-\frac{2}{\bar{\alpha}}
\big[(\frac{r}{r_{s}})^{\bar{\alpha}}-(\frac{r_{0}}{r_{s}})^{\bar{\alpha}}\big]}\, ,
\end{equation}
with the standard choices $r_{s}=20$ kpc, $\bar{\alpha}=0.17$ and 
$\rho_0=0.3$ GeV cm$^{-3}$. From this we can calculate the galactic
neutrino flux contribution as

\begin{equation}
\left(\frac{d\Phi_{\nu}}{dE_{\nu}}\right)_{\rm gal}=
\left( \frac{1.3\times 10^{-13}}{\rm cm^{2}\;sr\: s} \right)\,\frac{10^{28} \rm  \, s}{\tau_{Y}}
\,\frac{1 \, \rm PeV}{M_{Y}}\, \frac{1}{N}\frac{dN}{dE_{\nu}}\, .
\label{eq:galacticDMcontribution}
\end{equation}

The extra-galactic contribution to the neutrino flux~\cite{Ibarra:2007wg} 
is given by  

\begin{equation}
\frac{d\Phi_{\nu}}{dE_{\nu}}=\frac{\Omega_{DM}\, \rho_{c}}{4\pi\tau_{Y}M_{Y}H_{0}
\Omega_{M}^{1/2}}\int_{1}^{\infty}dy\frac{dN}{Nd(E_{\nu}y)}\frac{y^{-3/2}}
{\sqrt{1+(\Omega_{\Lambda}/\Omega_{M})y^{-3}}} \, ,
\end{equation}
where $y=1+z$, $z$ being the red-shift. This can also be written as 

\begin{eqnarray}
\left(\frac{d\Phi_{\nu}}{dE_{\nu}}\right)_{\rm ex-gal} & = &
\left(\frac{2.5\times10^{-13}}{\rm cm^{2} \, sr\, s}\right) \, \frac{10^{28}
  \, \rm s}{\tau_{Y}}\; \frac{1\,\rm PeV}{M_{Y}}\nonumber \\ & &
\times\int_{1}^{\infty}dy\frac{dN}{N \,
  d(E_{\nu}y)}\frac{y^{-3/2}}{\sqrt{1+(\Omega_{\Lambda}/\Omega_{M}) \,
    y^{-3}}}\, ,
\label{eq:extragalacticneutrino}
\end{eqnarray}
where the numerical values of the cosmological densities
$\Omega_{DM}=0.265$, $\Omega_{M}=0.315$,
$\Omega_{\Lambda}=0.685$, Hubble constant $H_0=67.3$ km/s/Mpc and 
$\rho_{c}=1.054\times10^{-5} h^2$ GeV/cm$^{3}$ were taken from 
\cite{Ade:2013zuv,Agashe:2014kda}. Here we want to stress that although in the fit 
we have fixed $\Omega_{DM}=0.265$ to be all the DM
density, in principle the LLP can constitute only part 
of the DM. The lifetime of the LLP $\tau_Y$ obtained
from the fit will have to be multiplied by a factor of $\kappa$
if the LLP constitute only a fraction $\kappa$ to the DM density.
Hence $\tau_Y$ obtained from the fit is the upper bound 
on the lifetime of $Y$. We refer the reader to Section~\ref{subsec:LLP_DM}
for a discussion on the LLP abundance.
Again the term $(1/N)dN/dE_{\nu}$ is the normalized 
neutrino energy spectrum calculated using PYTHIA in accordance to the 
LLP decay modes of a particular scenario.
In the above $M_Y$ and $\tau_Y$ are parameters to be fit to the data.

The cosmic unknown neutrino source contribution was estimated as a
power-law similar to \cite{Chen:2008yi,Chen:2013dza}  but using the 
convenient parametrization 

\begin{equation}
\left(\frac{d\Phi_{\nu}}{dE_{\nu}}\right)_{\rm
  pl}=\frac{C_0}{10^{8}}\times \frac{1}{E_{\nu}^2} \times \left(
\frac{E_{\nu}}{100 \; \rm TeV}\right)^{2-s}\, ,
\label{eq:diffuseneutrinos}
\end{equation}
where $C_0$ is the per-flavour normalization (1:1:1) and $s$ is the spectral index. 
These are parameters to be fit to the experimental observations. 
The total number of neutrino events expected in the $n$-th 
energy bin of IceCube, $N(E_n)$,  is calculated as

\begin{equation}
N(E_n)=T \times \Omega \times \sum_{j,\alpha} \, \int_{E_n}^{E_{n+1}} dE_\nu \, A^{\alpha}_{\rm
  eff}(E_\nu) \; \left( \frac{d\Phi_{\nu}}{dE_{\nu}} \right)^{\alpha}_j\, ,
\label{eq:neutrinorate}
\end{equation}
where $T$ is the exposure time, here 988 days~\cite{Aartsen:2014gkd}, 
$\Omega = 4\pi$ is the solid angle of coverage,
$A^{\alpha}_{\rm eff} (E_\nu)$ is the effective area for the neutrino flavour 
$\alpha$ taken from \cite{Aartsen:2013jdh}, 
$E_n$ and $E_{n+1}$ are the lower and upper energy limits of the bin 
and the sum $\sum_{j,\alpha}$ is performed over each lepton flavour $\alpha$ 
and the different contributions to the neutrino flux (e.g. Eqs. \eqref{eq:galacticDMcontribution},  
\eqref{eq:extragalacticneutrino}, \eqref{eq:diffuseneutrinos} and the atmospheric 
background neutrinos discussed in the next paragraph), labelled by $j$, 
for each scenario as shown in Table \ref{tab:Summary-of-Pvalues}. 
For scenarios that contain two decaying channels the 
branching ratio $r_{\nu N}$ is included in the computation of $(1/N)dN/dE_{\nu}$.

Finally we will also take into account the background neutrinos that arise 
from cosmic ray air showers, mainly from muon, $\pi/K$ and charm decays, 
simply by using the digitalized numbers for the atmospheric background 
from the IceCube paper~\cite{Aartsen:2014gkd}. 
We have extrapolated this background to higher energy bins (up to 10 PeV).

\section{Fitting the IceCube Data in Three Scenarios}
\label{sec:fitting}

We compare three scenarios in fitting the three years of IceCube
data~\cite{Aartsen:2014gkd}: (a) a single power-law, (b) a power-law
and a two-body LLP decay and (c) pure LLP decays according to the LLP
model described in Section~\ref{sec:model}.
In the analysis, we consider the energy domain from about 10 TeV up 
to 10 PeV, altogether 14 bins as shown for e.g. in Figure \ref{fig:powerlaw1}. 
Although the last three bins (from about 2 PeV to 10 PeV) have null 
events, they are important to help to disentangle the various hypotheses 
as we will show in Section~\ref{sec:future}.

In the scenario (a) we have two free parameters, $s$ and $C_0$
(Eq.~\ref{eq:diffuseneutrinos}), in the scenario (b) we have an extra
free parameter, the lifetime $\tau_Y$ while the LLP mass $M_Y$ is
fixed to be either $2.2$ or $4.0$ PeV and the branching ratio $r_{\nu N}={\rm
  BR}(Y \to \nu N)=1$.  Finally, in the scenario (c) we will also have
two free parameters $\tau_Y$ and $r_{\nu N}$
while $M_Y$ will also be fixed to be either $2.2$ or $4.0$ PeV.

In order to estimate the best fit values of the parameters in each
case we use the method of maximum likelihood by constructing a 
probability distribution function (pdf). We also  compute the p-value 
associated with each hypothesis $\rm H_0$. The description of our statistical 
procedure can be found in  Appendix~\ref{details}. 

The summary of best fit parameters and the corresponding p-values for
the hypothesis considered in this work are given in
Table~\ref{tab:Summary-of-Pvalues}.

\begin{table}
\begin{centering}
\begin{tabular}{|c|c|c|c|c|c|c|c|c|}
\hline 
$\rm H_{0}$ & $M_{Y}$ {[}PeV{]} & Scenario & $s$ & $C_{0}$ & $\tau_{Y}$ $\times 10^{28}$ {[}s{]} & $r_{\nu N}$ & $\chi_{\rm min}^{2}$ & $p$\tabularnewline
\hline 
\hline 
I & - & PL & $2.3$ & $0.6$  & - & - & $39.41$ & $0.5$\tabularnewline
\hline 
II.a & $2.2$ & $\mbox{PL}+\nu N$ & $2.43$ & $0.51$ & $5.26$ & - & $38.07$ & $0.45$\tabularnewline
\hline 
II.b & $4.0$ & $\mbox{PL}+\nu N$ & $2.76$ & $0.52$ & $2.72$ & - & $36.67$ & $0.58$\tabularnewline
\hline 
III.a & $2.2$ & $\nu N+4h$ & - & - & $0.73$ & $0.14$ & $42.53$ & $0.06$\tabularnewline
\hline 
III.b & $4.0$ & $\nu N+4h$ & - & - & $0.88$ & $0.35$ & $36.6$ & $0.56$\tabularnewline
\hline 
IV.a & $2.2$ & $\nu N+2h$ & - & - & $1.81$ & $0.56$ & $44.87$ & $0.01$\tabularnewline
\hline 
IV.b & $4.0$ & $\nu N+2h$ & - & - & $1.13$ & $0.23$ & $36.25$ & $0.57$\tabularnewline
\hline 
V & $4.0$ &  $\nu N$ & - & - & $1.9$ & - & $38.64$ & $0.24$\tabularnewline
\hline 
\end{tabular}
\par\end{centering}
\caption{\label{tab:Summary-of-Pvalues}Summary of best fit parameters
  and p-values for the hypothesis $\rm H_0$ considered in this work. PL
  stands for power-law, $C_0$ is given in GeV cm$^{-2}$ s$^{-1}$
  sr$^{-1}$ and $r_{\nu N}$ is the branching ratio of the
  channel $Y\rightarrow \nu N$.}
\end{table}

\subsection{Power-Law Fit ($\rm H_0= I$)}

\begin{figure}[htb]
\begin{center}
\includegraphics[scale=0.7]{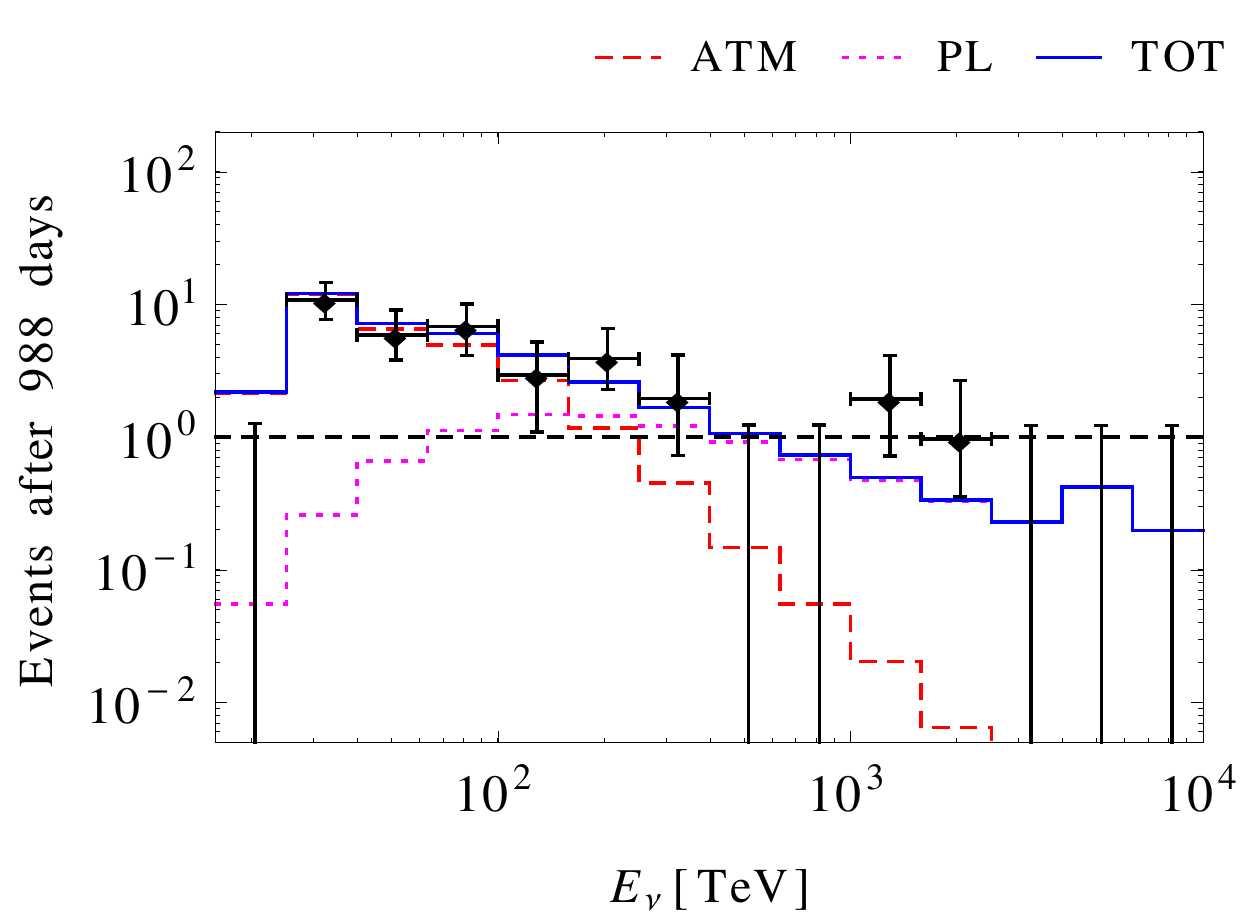}
\vspace{-2mm}
\end{center}
\vspace{-4mm}
\caption{Best fit curve for an unbroken power-law spectrum with
  $s=2.3$ and $C_0=0.6$ GeV cm$^{-2}$ s$^{-1}$ sr$^{-1}$. The
  IceCube data points (black crosses) are shown as well as the
  contributions from atmospheric background (ATM, red), the single
  power-law spectrum (PL, magenta) and the total contribution (TOT,
  blue).}
\label{fig:powerlaw1}
\end{figure}

In Figure~\ref{fig:powerlaw1} we show the best fit curve for an
unbroken power-law spectrum fit to the data. Although it is at this
point unclear if this power-law behaviour can be explained by a single
astrophysical type of source, this hypothesis is the simplest one.  We
show the atmospheric background contribution (red curve), 
the power-law contribution (magenta curve) as well as
the sum of the atmospheric background and the power-law signal fit
(blue curve) for our best fit value at $s=2.3$ and $C_0=0.6$ GeV
cm$^{-2}$ s$^{-1}$ sr$^{-1}$, which corresponds to a p-value of $0.5$.
From this we see that the power-law contribution is only significant 
for energies $\gsim$100 TeV, the lower energy part of the spectrum 
is dominated by the atmospheric background. If this hypothesis is true 
there should be events in the gap between 400 TeV and 1 PeV and 
above 2 PeV in the future. 

An unbroken power-law  spectrum such as this one may arise from optically thin 
galactic neutrino sources~\cite{Anchordoqui:2013qsi}.
It has been pointed out that cosmic ray interactions with gas, such as expected 
around supernova remnants, seem to be able to produce smooth 
neutrino spectra~\cite{Loeb:2006tw}.

\begin{figure}[htb]
\begin{center}
\includegraphics[scale=0.9]{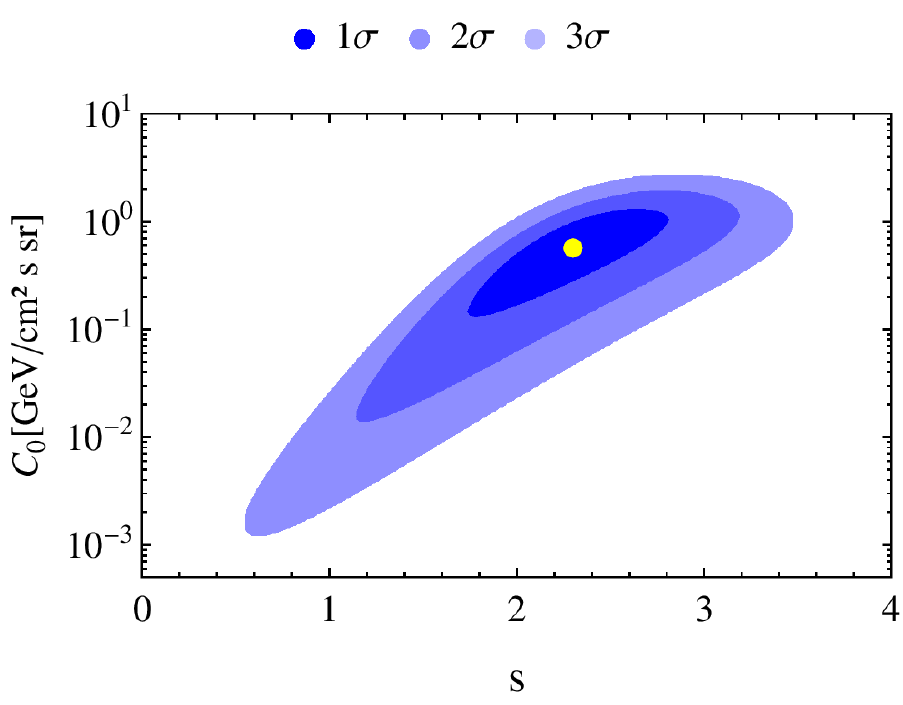}
\vspace{-2mm}
\end{center}
\vspace{-4mm}
\caption{Contour plot of the allowed region in the plane $C_0 \times s$
 at 1, 2 and 3$\sigma$ CL for the power-law hypothesis.}
\label{fig:plflux}
\end{figure}

Here a note is in order. Our best-fit value is compatible to IceCube
spectral index fit~\cite{Aartsen:2014gkd} within 1$\sigma$ since at
this confidence level $1.72 \le s \le 2.83$.  We show in Figure~\ref{fig:plflux}
the correlation between $C_0$ and the spectral index $s$ for our
fit. We can see that a small change in $s$ can cause a significant
change in $C_0$ and vice-versa, so the best fit values of these
parameters are not at this point very significant.

Also we would like to comment on the new IceCube veto-based technique
developed to study the neutrino spectrum between 10 and 100
TeV~\cite{Aartsen:2014muf}. Using this new method they were able to
better understand their background at lower energies.  Although this
could have an impact on the best fit values of the parameters in our
analysis, we do not believe our conclusions would change.  For
completeness we present in Table~\ref{tab:Confidence-level-intervals}
in Appendix~\ref{summary} the 1, 2 and 3$\sigma$ confidence intervals for $s$ and 
$C_0$ for our fit.

\subsection{Power-Law + Long-Lived Particle Two Body Decay Fit ($\rm H_0 = II$)}

\begin{figure}[htb]
\begin{center}
\includegraphics[scale=0.55]{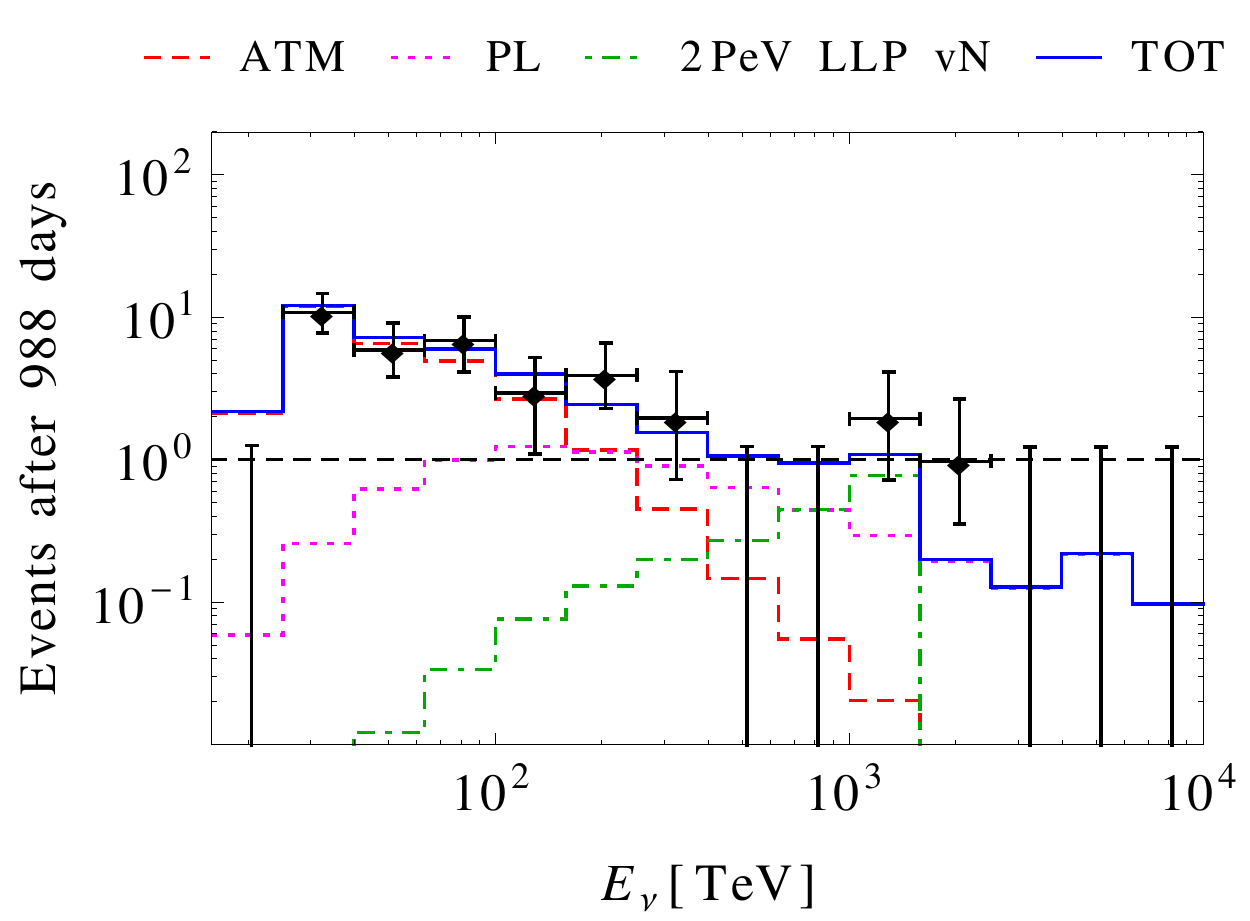}
\includegraphics[scale=0.55]{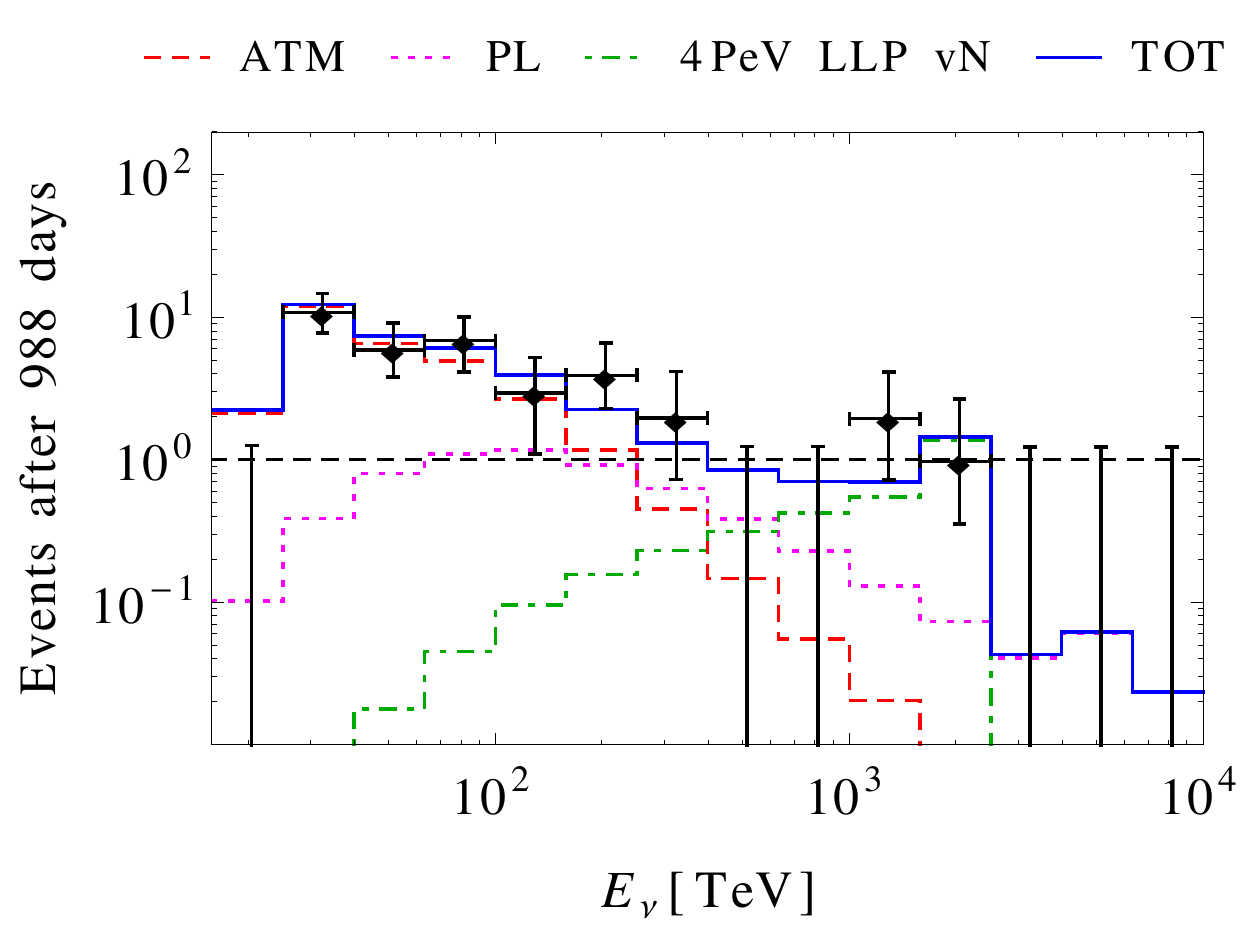}
\vspace{-2mm}
\end{center}
\vspace{-4mm}
\caption{Best fit curve for an unbroken power-law spectrum combined
  with the LLP decay $Y\to \nu N$. The left panel is for $M_Y=2.2$
  PeV, $s=2.43$, $C_0=0.51$ GeV cm$^{-2}$ sr$^{-1}$ s$^{-1}$ and
  $\tau_Y=5.26 \times 10^{28}$ s. The right panel is for $M_Y=4$ PeV,
  $s=2.76$, $C_0=0.52$ GeV cm$^{-2}$ sr$^{-1}$ s$^{-1}$ and
  $\tau_Y=2.72 \times 10^{28}$ s. The IceCube data points (black
  crosses) are shown as well as the contributions from atmospheric
  background (ATM, red), single power-law spectrum (PL, magenta), 
  LLP decay (LLP, green) and  the total contribution (TOT, blue).}
\label{fig:pl2-4TeV}
\end{figure}

In Figure~\ref{fig:pl2-4TeV} we show the best fit curve for a fit of
the data with a contribution from a power-law spectrum combined with a
contribution from the LLP decay $Y\to \nu N$. On the left panel we
show the case $M_Y=2.2$ PeV, for the best fit $s=2.43$, $C_0=0.51$ GeV
cm$^{-2}$ sr$^{-1}$ s$^{-1}$ and $\tau_Y=5.26 \times 10^{28}$ s,
corresponding to a p-value of $0.45$. On the right panel we show the
case $M_Y=4$ PeV, for the best fit $s=2.76$, $C_0=0.52$ GeV cm$^{-2}$
sr$^{-1}$ s$^{-1}$ and $\tau_Y=2.72 \times 10^{28}$ s, corresponding
to a p-value of $0.58$.  The hypotheses I, II.a and II.b all have very
similar p-values and at this point seem to be indistinguishable.  In
both cases II.a and II.b, the power-law contribution is similar to the
single power-law fit and the LLP decay basically contributes to the 1
or 2 PeV energy bin, depending on $M_Y$. The future content of these
bins can help to distinguish the hybrid hypothesis from the single
power-law one.
 
The confidence intervals for the fitted parameters can also be found
in Table~\ref{tab:Confidence-level-intervals} in Appendix~\ref{summary}. 
From these intervals we see that if one allows for both a LLP two-body
decay as well as a contribution from a power-law spectrum, there is a
minimum lifetime for the LLP compatible with the data, but longer
lifetimes are clearly also possible because in this case in practice
we revert back to the single power-law case.

\subsection{Pure Long-Lived Particle Decays Fit ($\rm H_0=III,IV$ and $\rm V$)}

We have examined  three scenarios for pure LLP decays. For $\rm H_0=III$ we 
have the two comparable decay modes: $Y\to \nu N$  and $Y\to 4 h$.
For $\rm H_0=IV$ we have two comparable two-body decay modes: 
$Y\to \nu N$  and $Y\to 2 h$, whereas for $\rm H_0=V$ we have only a single 
decay contribution $Y\to \nu N$.

\begin{figure}[htb]
\begin{center}
\includegraphics[scale=0.55]{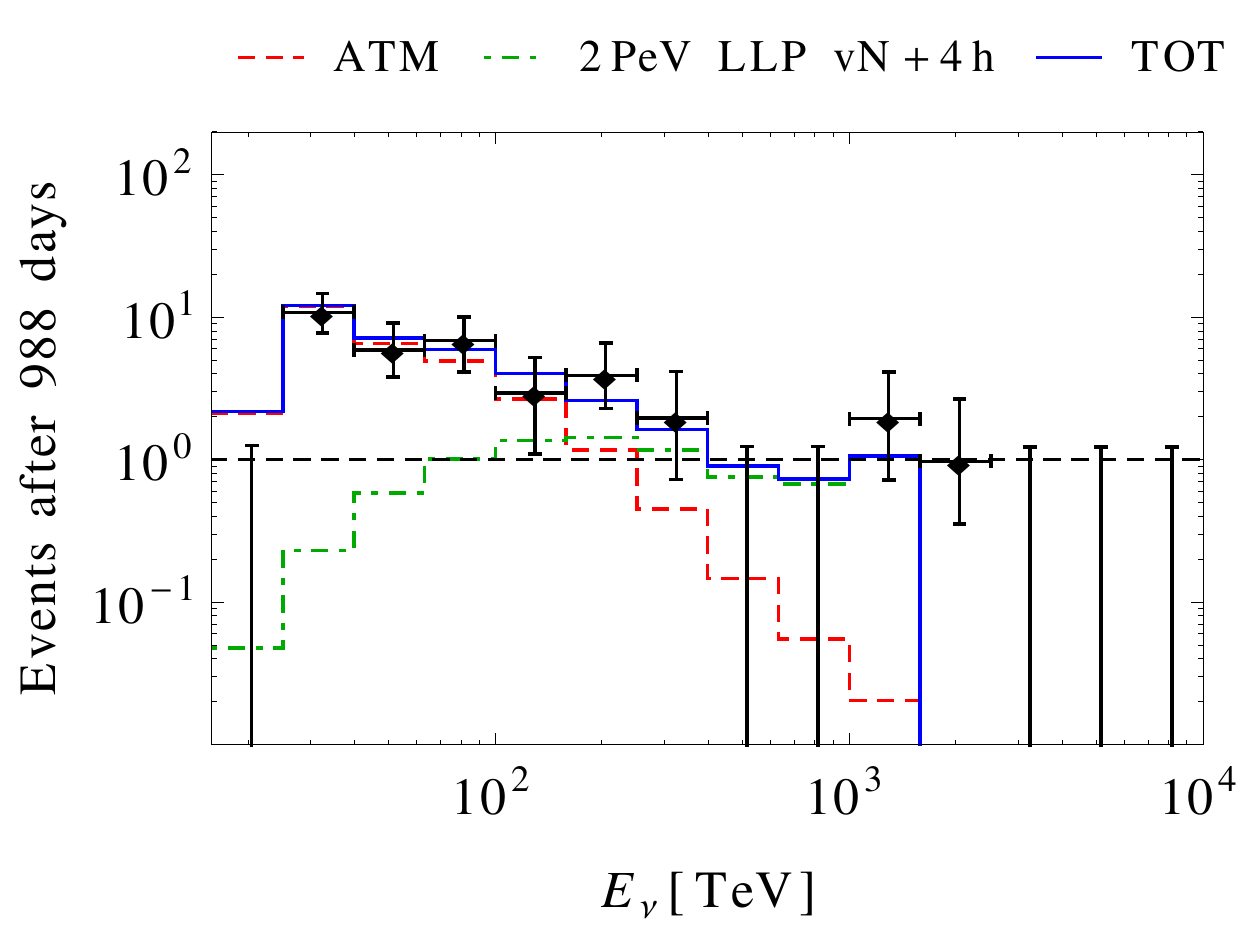}
\includegraphics[scale=0.55]{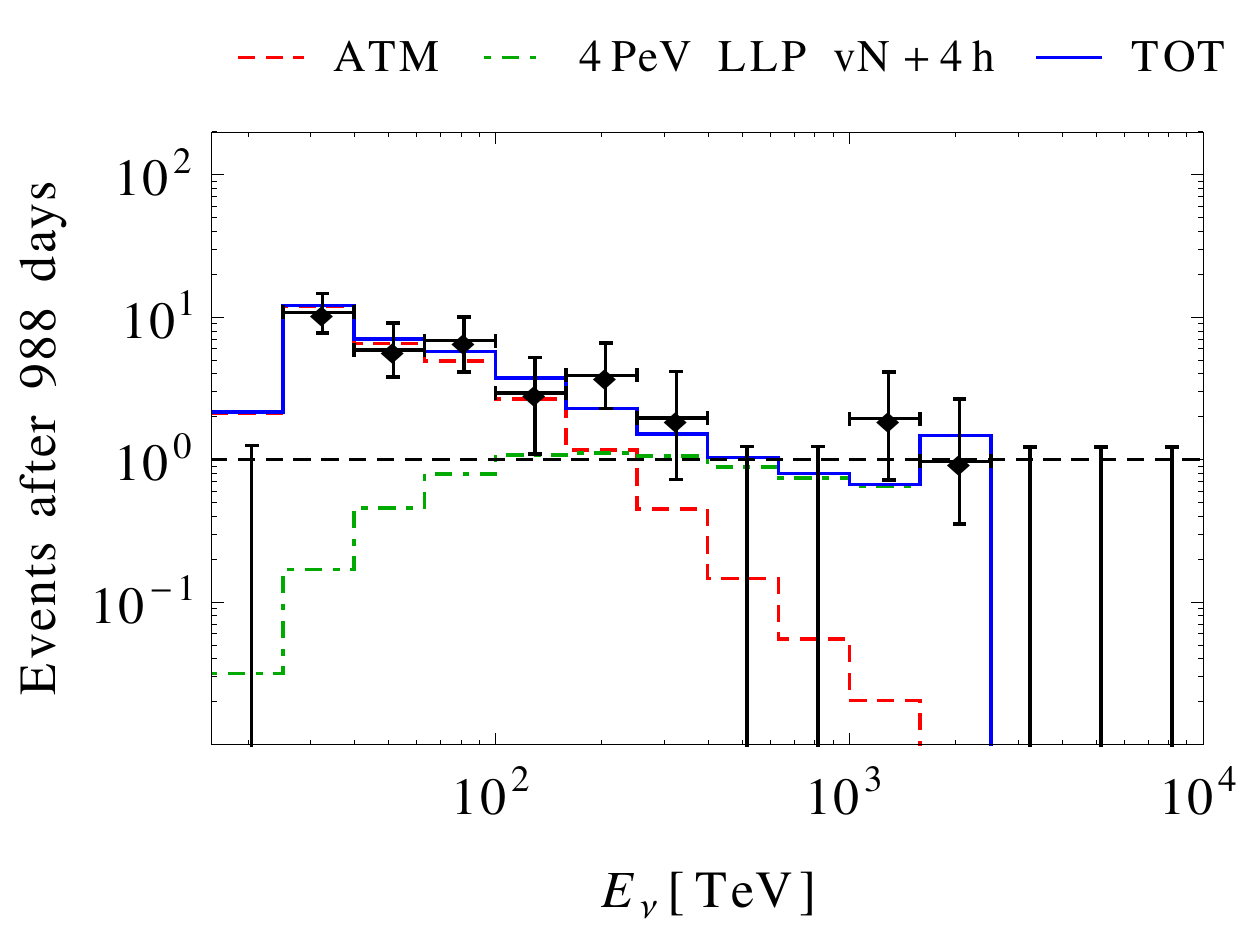}
\vspace{-2mm}
\end{center}
\vspace{-4mm}
\caption{Best fit curve for pure LLP decays into $Y\to \nu N$ and $Y
  \to 4h$. The left panel is for $M_Y=2.2$ PeV, $\tau_Y=0.73 \times
  10^{28}$ s and $r_{\nu N}=$ 14\%.  The right panel is for
  $M_Y=4$ PeV, $\tau_Y=0.88 \times 10^{28}$ s and $r_{\nu N}=$
  35\%.  The IceCube data points (black crosses) are shown as well as
  the contributions from atmospheric background (ATM, red), LLP decays
  (LLP, green) and the total contribution (TOT, blue).}
\label{fig:llp1}
\end{figure}

In Figure~\ref{fig:llp1} we show the best fit curve for pure LLP decay
into the two modes $Y \to \nu N$ and $Y\to 4 h$.  On the
left panel we show the case $M_Y=2.2$ PeV, for the best fit 
$\tau_Y=0.73\times 10^{28}$ s and $r_{\nu N}=0.14$, corresponding to a
p-value of $0.06$.  On the right panel we show the case $M_Y=4$ PeV, for
the best fit $\tau_Y=0.88 \times 10^{28}$ s and $r_{\nu N}=0.35$,
corresponding to a p-value of $0.56$.

\begin{figure}[htb]
\begin{center}
\includegraphics[scale=0.55]{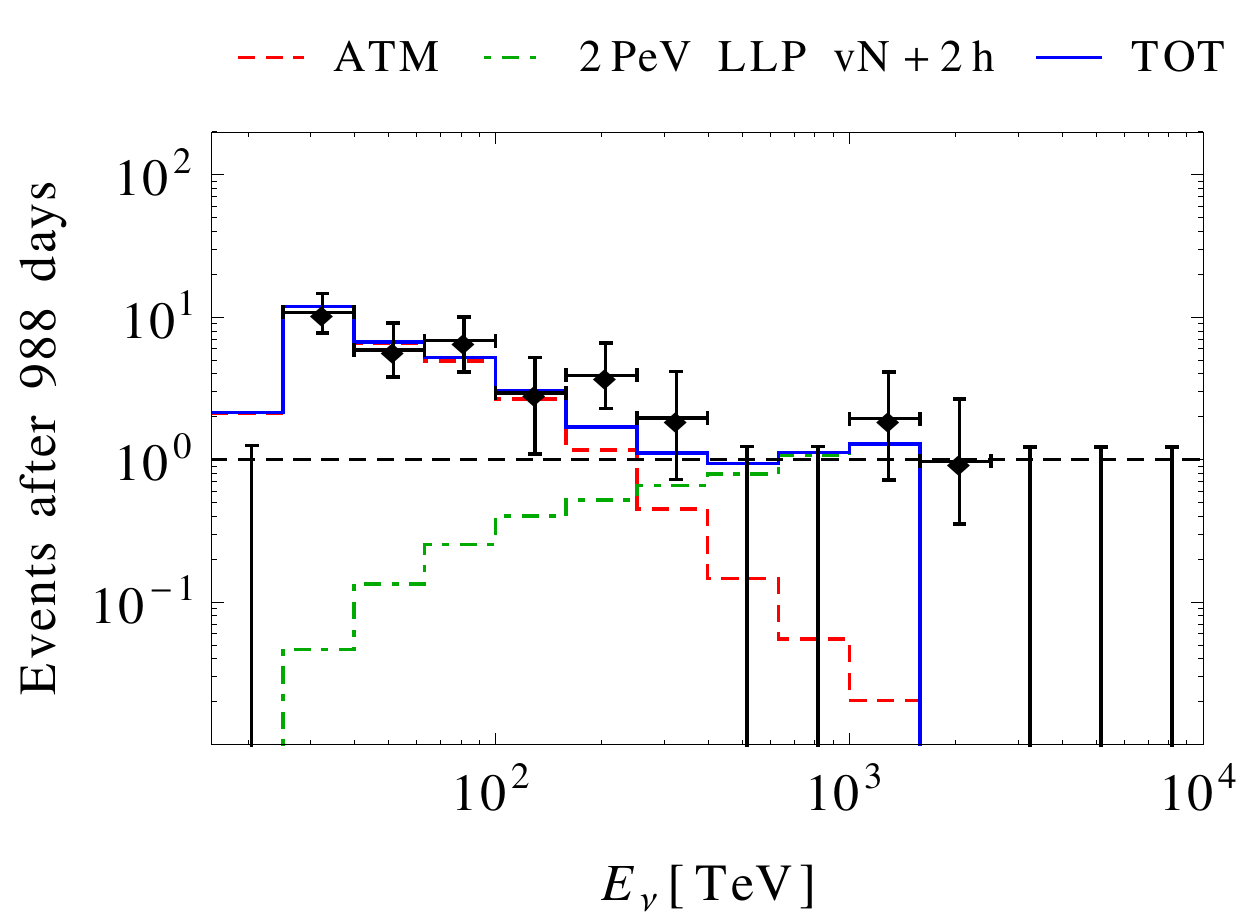}
\includegraphics[scale=0.55]{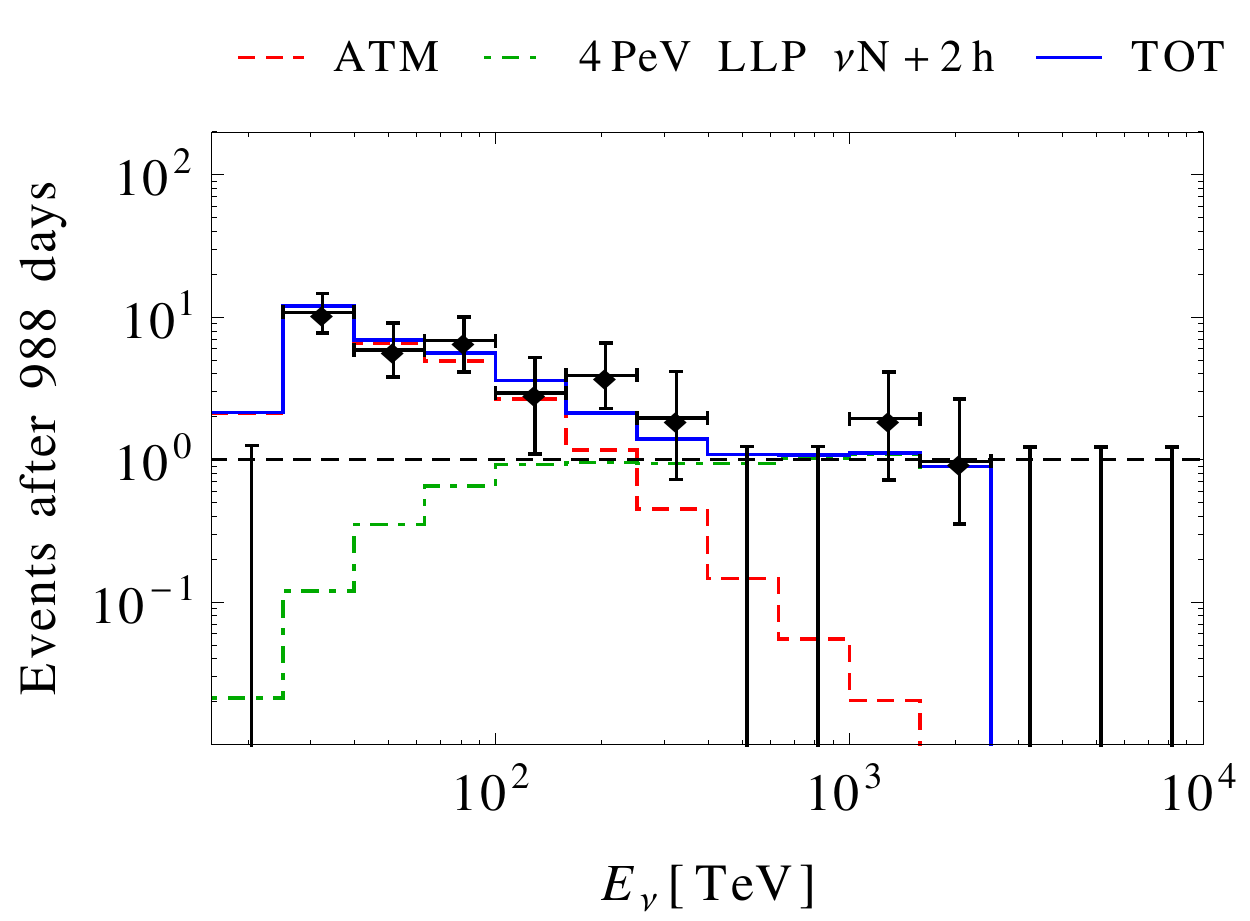}
\vspace{-2mm}
\end{center}
\vspace{-4mm}
\caption{Best fit curve for pure LLP decays into $Y\to \nu N$ and $Y
  \to 2h$. The left panel is for $M_Y=2.2$ PeV, $\tau_Y=1.81 \times
  10^{28}$ s and $r_{\nu N}=$ 56\%.  The right panel is for
  $M_Y=4$ PeV, $\tau_Y=1.13 \times 10^{28}$ s and $r_{\nu N}=$
  23\%.  The IceCube data points (black crosses) are shown as well as
  the contributions from atmospheric background (ATM, red), LLP decays
  (LLP, green) and the total contribution (TOT, blue).}
\label{fig:llp2}
\end{figure}

In Figure~\ref{fig:llp2} we show the best fit curve for pure LLP decay
into the two modes $Y \to \nu N$ and $Y\to 2 h$.  On the
left panel we show the case $M_Y=2.2$ PeV, for the best fit 
$\tau_Y=1.81\times 10^{28}$ s and $r_{\nu N}=0.56$, corresponding to a
p-value of $0.01$.  On the right panel we show the case $M_Y=4$ PeV, for
the best fit $\tau_Y=1.13 \times 10^{28}$ s and $r_{\nu N}=0.23$,
corresponding to a p-value $0.57$.  

Clearly the cases with $M_Y=$ 2.2 PeV are very unfavourable, but the
cases with $M_Y=$ 4 PeV are consistent with data.  In fact, the fit
with $M_Y=$ 4 PeV and the LLP decaying into $\nu N$ and $2 h$ presents
one of the highest p-value of all cases studied.  We see that the LLP decays
start to contribute to the spectrum at energies $\gsim 70$ TeV up to 2
PeV, so these scenarios predicts a sharp cutoff in the spectrum above 2
PeV that could be confirmed by future data.

\begin{figure}[htb]
\begin{center}
\includegraphics[scale=0.8]{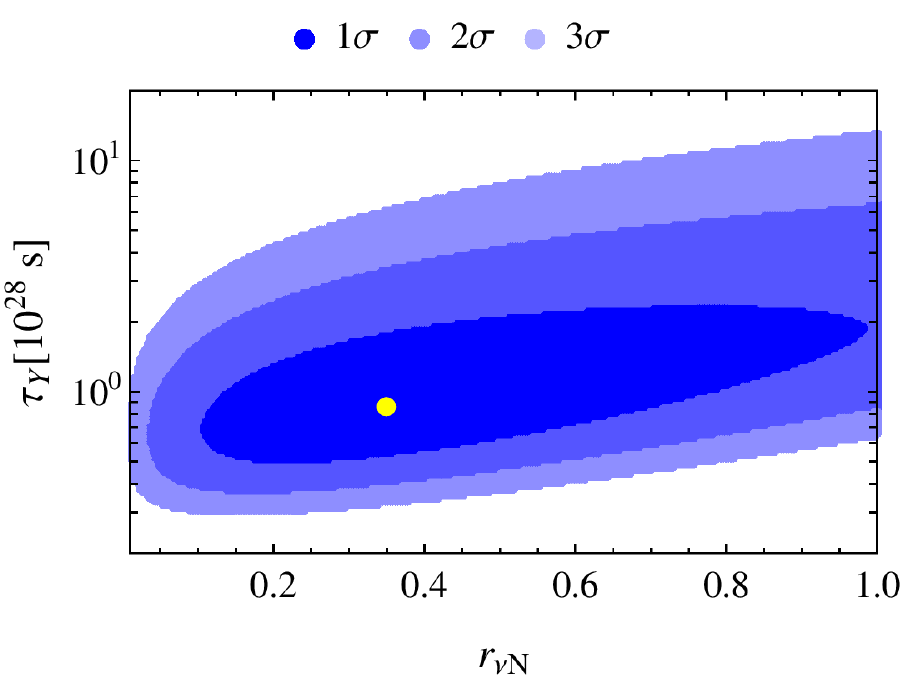}
\includegraphics[scale=0.8]{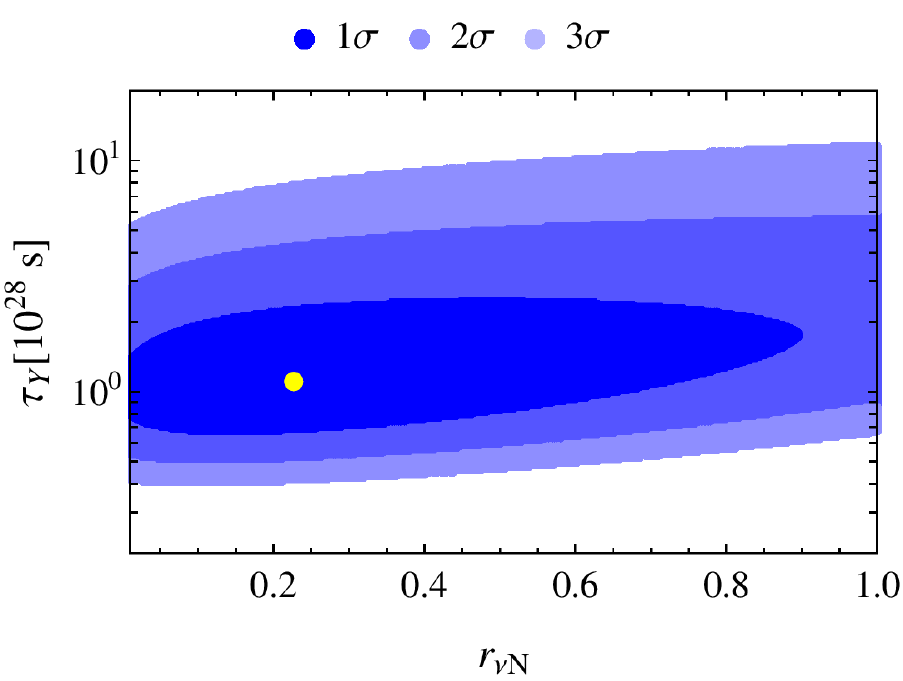}
\vspace{-2mm}
\end{center}
\vspace{-4mm}
\caption{Contour plots of the allowed regions in the plane $\tau_Y
  \times r_{\nu N}$ at 1, 2 and 3 $\sigma$ CL for the hypotheses III.b
  (left panel) and IV.b (right panel). }
\label{fig:llpflux}
\end{figure}

In Figure~\ref{fig:llpflux}  we show the correlation between $\tau_Y$ and 
the branching ratio $r_{\nu N}$ for $M_Y=$ 4 PeV and hypotheses III and 
IV. We see the $Y$ lifetime is slightly correlated with $r_{\nu N}$, 
more so for III than for IV. Also  when we have $Y \to 2h$ it is 
possible, even at 1$\sigma$, to have $r_{\nu N}=0$ while when
we have $Y \to 4 h$ we need at least some $Y\to \nu N$ contribution in order 
to explain the data.

\begin{figure}[htb]
\begin{center}
\includegraphics[scale=0.7]{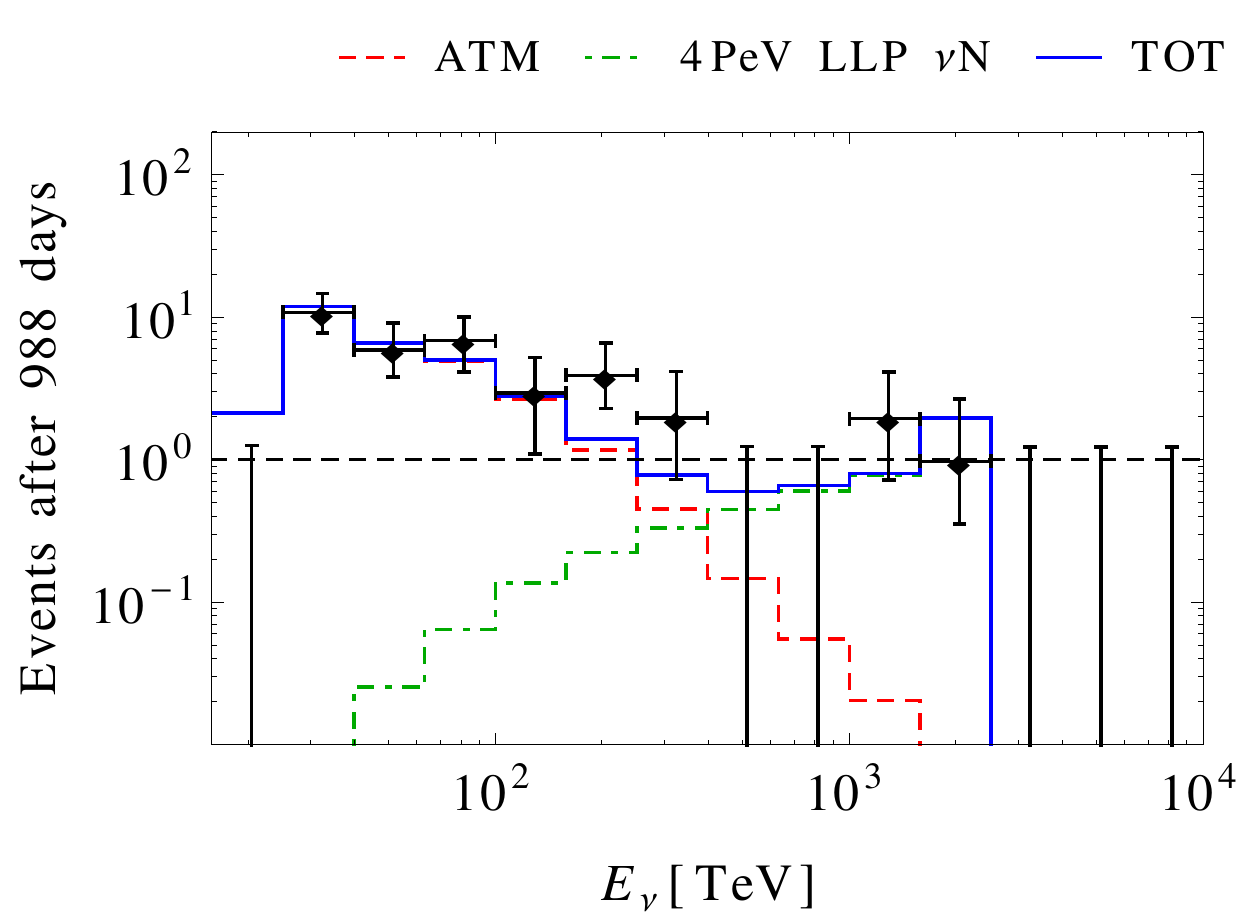}
\vspace{-2mm}
\end{center}
\vspace{-4mm}
\caption{Best fit curve for a $M_Y=$ 4 PeV LLP decaying into $Y\to \nu
  N$ with $\tau_Y=1.9 \times 10^{28}$ s. 
The IceCube data points (black crosses) are shown as well as
  the contributions from atmospheric background (ATM, red), the LLP decay
  (LLP, green) and the total contribution (TOT, blue).}
\label{fig:llp3}
\end{figure}

Finally, we also investigate the single decay mode $Y\to \nu N$ for
$M_Y=4$ PeV for which we get the best fit $\tau_Y=1.9\times 10^{28}$ s 
and a corresponding p-value of $0.25$. This case is shown in Figure~\ref{fig:llp3}. 
We see that in this case instead of just a peak at 2 PeV, 
there is a cascade tail of lower energies neutrinos due to EW corrections
(partially also due to the extragalactic neutrinos from LLP decay 
that redshifts to lower energies).
Nevertheless the LLP decay only starts to contribute at much higher energies, 
around 500 TeV, so up to that point the spectrum has to be entirely explained by the
atmospheric background.  Also this scenario predicts a cutoff in the
spectrum after 2 PeV. Just by looking at Figure~\ref{fig:llp3} we see
that the data in the two bins bellow 500 TeV and the bin at 1 PeV
seem to be higher than the theoretical prediction. You can find in
Table~\ref{tab:Confidence-level-intervals} in Appendix~\ref{summary} 
the allowed intervals of the parameters for all cases considered here.

To conclude this section we note that at this point the data seems to
be equally compatible with a single power-law spectrum, a power-law plus
$Y\to \nu N$ spectrum or a spectrum due to a 4 PeV LLP decaying into
$\nu N$ and $2h$ or $4h$.  

\subsection{Constraints from Gamma-ray and Antiproton Observations}

We now briefly discuss the question of whether our LLP decay scenario
for IceCube high-energy events is consistent with the limits imposed
by diffuse gamma-ray and antiproton observations. Here, we limit
ourselves to a sketchy description by just reviewing the results in
the existing literatures.\footnote{
We note that the limit derived for super-heavy DM applies to our LLP 
scenario because the mass density of LLP cannot exceeds that of DM. 
That is, the DM mass density gives a maximum possible value of 
LLP mass density, and hence if DM evades a limit our LLP does. }
%
As discussed by the authors of \cite{Murase:2012xs}, the cascade
gamma-ray bound is largely DM mass-independent at sufficiently high
masses, because it is essentially bolometric in nature. It allows us
relatively DM mass independent conclusion. Also the gamma-ray limits
at very high masses is weaker than the limit for neutrinos which was
obtained \cite{Murase:2012xs} assuming non observation of three years
run of IceCube. When applied to our case, it means that models of LLP
decay which explains IceCube neutrino excess would be free from the
diffuse gamma ray bound, even though the new Fermi-LAT data at higher 
energies \cite{Ackermann:2014usa} makes the consistency more nontrivial \cite{murase}.
The authors of \cite{Esmaili:2014rma} also reached the similar conclusion 
on gamma ray bound with the recent Fermi data.

More specifically in our case, the model we examined in this section
is much safer than the generic LLP decay scenario, because the decay
products, neutrinos, gammas, and electrons, etc. from Higgs boson is
about 10 times less prominent compared to those from $b \bar{b}$ at
low energies (see also \cite{Ahlers:2013xia}).
It appears that the PAMELA antiproton limit is also cleared by our LLP
scenario, as one can see in Figure~6 in \cite{Garny:2012vt}. By
extrapolating three-orders of magnitude in the DM mass from those in
Figure~6 the lifetime lower bound is well below $10^{27}$ s.

\section{Future Data Perspectives}
\label{sec:future}

In view of the fact that at this point the data seems to be compatible
with many different cases we would like to establish the necessary
exposure time in order to exclude a given hypothesis $\rm H_j$ assuming
the true explanation of the data is $\rm H_i$. We refer the reader to
Appendix~\ref{details} for details of the statistical calculation.
Our discussion in this section relies only on energy spectrum 
information.\footnote{There may be other ways to distinguish the various hypotheses including ours,
for example, by pinning down the sources, possible identification of galactic-extragalactic
components, and by correlating with gamma ray observation.}

\begin{figure}[ht]
\begin{centering}
\includegraphics[scale=0.7]{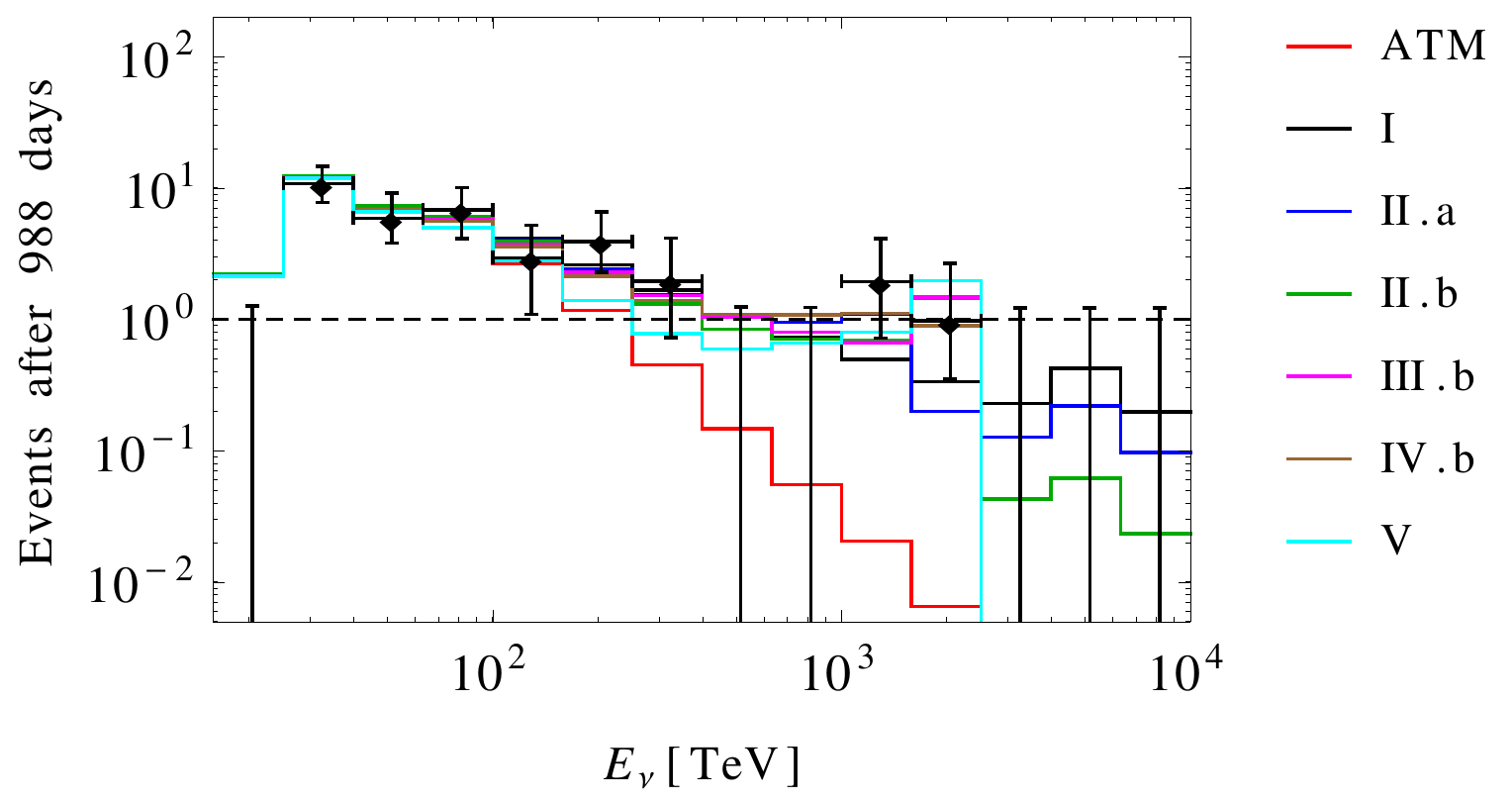}
\par\end{centering}
\caption{\label{fig:Extrapolation-of-the-models} The best fit for all the
  hypotheses with the current data and the atmospheric background 
  extrapolated up to the energy range [10\, \rm
    TeV,\,10\, \rm PeV].  
}
\end{figure}

In Figure~\ref{fig:Extrapolation-of-the-models} we show the curves for
all the hypotheses best fit with the current data. 
From this we can clearly see that data at higher energies 
will be able to help to disentangle the various hypotheses.

In Table \ref{tab:Exclusion-Time.} we show the results of our
computations for all possible combinations of the hypotheses we consider. 
In the columns we place the true hypothesis $H_{i}$ while
in the rows we place the hypothesis to be excluded $H_{j}$. 
We define the exclusion time as $T=f_{T}\times 988$ days.  
In each cell we write the vector $(f_{T},{\rm p}\times100)$ in order to indicate the
necessary exposure time and the corresponding p-value as a percentage
computed at that time. Also, we use the notation $(f_{T},{\rm p}\times100)^{*}$ to
identify combinations that can be distinguish but in a very long time
and $(f_{T},{\rm p}\times100)^{**}$ for those combinations that cannot be
distinguished at any time.

\begin{table}[htb]
\begin{centering}
\begin{tabular}{|c|c||c|c|c|c|c|c|}
\cline{2-8} 
\multicolumn{1}{c|}{} & \multicolumn{2}{c|}{I} & II.a & II.b & III.b & IV.b & V\tabularnewline
\hline 
I & \multicolumn{2}{c|}{-} & $(25,3)^{*}$ & $(8,3)$ & $(8,2)$ & $(9,2)$ & $(4.5,3)$\tabularnewline
\hline 
II.a & \multicolumn{2}{c|}{$(100,100)^{**}$} & - & $(7,3)$ & $(8,2)$ & $(12,5)$ & $(4.5,2)$\tabularnewline
\hline 
II.b & \multicolumn{2}{c|}{$(100,100)^{**}$} & $(23,3)^{*}$ & - & $(75,4)^{*}$ & $(25,4)^{*}$ & $(100,100)^{**}$\tabularnewline
\hline 
III.b & \multicolumn{2}{c|}{$(4.5,2)$} & $(6,3)$ & $(34,1)^{*}$ & - & $(28,5)^{*}$ & $(100,100)^{**}$\tabularnewline
\hline 
IV.b & \multicolumn{2}{c|}{$(4.5,1)$} & $(9,4)$ & $(20,4)^{*}$ & $(45,5)^{*}$ & - & $(100,100)^{**}$\tabularnewline
\hline 
V & \multicolumn{2}{c|}{$(2.5,1)$} & $(3.5,4)$ & $(12,4)$ & $(11,4)$ & $(9,3)$ & -\tabularnewline
\hline 
\end{tabular}
\par\end{centering}
\caption{\label{tab:Exclusion-Time.}Estimation of the exclusion time needed to 
eliminate a hypothesis. In each cell we write
the vector $(f_{T},{\rm p}\times100)$. We use the notation $(f_{T},{\rm p}\times100)^{*}$
to identify combinations that can be excluded but in a very long
time and $(f_{T},{\rm p}\times100)^{**}$ for those combinations that cannot
be excluded.}
\end{table}

We see that assuming a power-law spectrum, the solely LLP decay models can 
be excluded with two to five times the current data due to the absence of neutrino
events beyond the cutoff energy. If we introduce a LLP component to the power-law
it will take a longer time to distinguish from the solely LLP decay models because
in this case the power-law spectral index turns out to be larger and predicts less events
beyond the cutoff. A single power-law and a power-law plus the LLP decay is rather 
difficult to distinguish unless one constraints the lifetime of the LLP to a maximum value. 
If the future data is consistent with solely LLP decays it will be difficult to
distinguish among the possible contributing decay mode scenarios, but
the sharp cutoff in the spectrum will certainly indicate a LLP decay component.

\section{A Model for a Long-Lived Particle}
\label{sec:model}

In this section we present a consistent model for a LLP that could give rise
to the decay channels used to fit the data in the previous sections. 
We must emphasize that although we rely on this particular model to fit the 
IceCube high-energy excess events, it is by no means the unique model that 
can explain the IceCube data with or without the power-law component. 
However, we hope that it serves as the existence proof of 
such models that can explain the data only by LLP decays. 

As seen in the previous sections, we can accommodate the IceCube data with 
the following decay channels: $Y \to \nu N$,  $Y\to 2h$ and $Y\to 4h$. 
To accomplish this, we introduce two complex scalar fields, $Y$ and $X$, 
that are singlets under the SM gauge group. For the fermionic sector, 
we introduce a new vectorlike pair of fermion doublets 
$\Psi_L = (\psi_L^0,\psi_L^-)^T$ and $\Psi_R = (\psi_R^0,\psi_R^-)^T$, 
and another right-handed fermion singlet $N_R$.\
This model has no SM gauge anomaly since  $\Psi_L$ and $\Psi_R$ 
are vectorlike under the SM gauge interactions while $N_R$ is a singlet.
In addition, we assume these new fields to be charged 
under a new $U(1)_X$ symmetry according to Table~\ref{tab1}. 
In the following we will consider both possibilities that 
$U(1)_X$ is global or local, keeping in mind that if $U(1)_X$
is gauged, we need to introduce $N_L$ to cancel the $U(1)_X$ gauge anomaly.

\begin{table}[htb]
\begin{center}
\begin{tabular}{|c|c|c|c|}
  \hline 
  New fields & $SU(2)_L$ & $U(1)_Y$ & $U(1)_X$   
  \\ 
  \hline 
  \phantom{$\Big|$}     $X$ &  1  &  0   &  1   \\
  \hline 
  \phantom{$\Big|$}     $Y$ &  1  &  0   & -2   \\
  \hline 
  \phantom{$\Big|$}$\Psi_L$ &  2  & -1/2 &  2   \\
  \hline 
  \phantom{$\Big|$}$\Psi_R$ &  2  & -1/2 &  2   \\
  \hline 
  \phantom{$\Big|$}$N_R$    &  1  &  0   &  2   \\
  \hline 
\end{tabular}
\end{center}
\caption{New fields of the model and their respective 
assignments under $SU(2)_L\times U(1)_Y\times U(1)_X$.}
\label{tab1}
\end{table}

\subsection{The Scalar Sector}

Now we will describe the scalar sector of our model with the following
scalar potential
\begin{eqnarray}
V\left(X,Y,H\right) & = & \frac{1}{4}\lambda_{X}\left(X^{\dagger}X-w^{2}\right)^{2}+\frac{1}{4}\lambda_{H}\left(H^{\dagger}H-v^{2}\right)^{2}\nonumber \\
 &  & +\frac{1}{4}\lambda_{Y}\left(Y^{\dagger}Y\right)^{2}+M_{Y}^{2}\; Y^{\dagger}Y\nonumber \\
 &  & +\lambda_{HX}\left(H^{\dagger}H-v^{2}\right)\left(X^{\dagger}X-w^{2}\right)+\lambda_{XY}\left(X^{\dagger}X-w^{2}\right)Y^{\dagger}Y\nonumber \\
 &  & +\lambda_{HY}\left(H^{\dagger}H-v^{2}\right)Y^{\dagger}Y
-\left( \mu_{XY} \; XXY+ {\rm H.c.} \right),
\end{eqnarray}
where $H$ is the SM Higgs doublet. We assume that all the
dimensionless couplings $\lambda$'s are positive. The complex
dimension one coupling $\mu_{XY}$ can be made real by redefining $X$
and/or $Y$ fields. Without loss of generality, in the following,
$\mu_{XY}$ is taken to be real and positive.  We further assume that
the new physics scale $\left<X\right> = w \gg v = 174 $ GeV.  As we
will see below, the LLP is approximately $Y_R$ (the real part of $Y$)
and hence $M_Y$ is at the PeV scale.  We also take $M_Y^2 > 0$ such
that no large vacuum expectation value (vev) will be induced for $Y$.  
Instead, a small vev for $Y$ is induced through the $\mu_{XY}$ term:
\begin{equation}
\left<Y\right> = u = \mu_{XY} w^2/M_Y^2.
\label{eq:Y_vev}
\end{equation}
where we have assumed the condition $\mu_{XY} w^2/M_Y^3 \ll 1$.  As we
will see later, this condition is always fulfilled due to the
longevity requirement of the LLP.  In fact if $\mu_{XY} \to 0$, there
will be a $Z_2$ symmetry such that: $\Psi_{L,R} \to -\Psi_{L,R}$ and
$Y \to -Y$ (see Eq.~\eqref{eq:fermion_lag}).  Hence $\mu_{XY}$
controls the lifetime of our LLP.  A small $\mu_{XY}$ is technically
natural since in the limit $\mu_{XY} \to 0$, there is an enhanced
symmetry $U(1)^2$ which corresponds to independent phase rotations of
$X$ and $Y$.

$U(1)_X$ is spontaneously broken when $X$ acquires a vev $w$.  If the
$U(1)_X$ is a global symmetry, we will have one massless
Nambu-Goldstone boson (NGB). According to Ref.~\cite{Chang:1984ip}, if
one considers $10^9 \, \rm{GeV} \lesssim$ $w$ $\lesssim 10^{12}$ GeV, the NGB
will decouple much before the neutrino decoupling temperature $T \sim
$ MeV and hence its temperature will red-shift to a much lower value
and have small contribution to the energy density at the time of
nucleosynthesis.  Assuming the SM degrees of freedom, it is possible
to have up to 37 NGBs.  Alternatively, $U(1)_X$ can be gauged and
instead of a NGB, we will have a massive gauge boson associated with
$U(1)_X$. In this case the $w$ scale can be relaxed to a lower value.

In the following we write $Y = u + (Y_R + i Y_I)/\sqrt{2}$, 
$X = w + (X_R + i X_I)/\sqrt{2}$, $H = v + h/\sqrt{2}$.
Then the mass matrices for the fields $\left\{h, Y_R, X_R\right\}$ 
and $\left\{ Y_I, X_I \right\}$ are respectively
\begin{eqnarray}
M_{R}^{2} & = & \left(\begin{array}{ccc}
\lambda_{H}v^{2}+\lambda_{HY}u^{2} & 2\lambda_{HY}uv & 2\lambda_{HX}vw\\
2\lambda_{HY}uv & M_{Y}^{2}+\frac{3}{2}\lambda_{Y}u^{2} & 2\left(\lambda_{XY}u-\mu_{XY}\right)w\\
2\lambda_{HX}vw & 2\left(\lambda_{XY}u-\mu_{XY}\right)w & \lambda_{X}w^{2}+\left(\lambda_{XY}u-2\mu_{XY}\right)u
\end{array}\right)\, ,
\end{eqnarray}
and
\begin{eqnarray}
M_{I}^{2} & = & \left(\begin{array}{cc}
M_{Y}^{2}+\frac{1}{2}\lambda_{Y}u^{2} & 2\mu_{XY}w\\
2\mu_{XY}w & \left(\lambda_{XY}u+2\mu_{XY}\right)u
\end{array}\right)\, .
\end{eqnarray}

The longevity of LLP requires tiny $\mu_{XY}$, 
which from Eq. \eqref{eq:Y_vev} implies small mixing 
between $Y_R $ and $X_R$, $h$ and $Y_R$,
and $Y_I$ and $X_I$. 
On the other hand, the mixing between $h$ and $X_R$ 
is controlled by the ratio
\begin{eqnarray}
\delta_{HX} & \equiv & \frac{4\lambda_{HX}^{2}}{\lambda_{H}\lambda_{X}}.
\end{eqnarray}
For the SM Higgs boson mass $M_{h}=125$ GeV the still allowed branching ratio 
for the Higgs invisible decay width is in the ball-park of 20~\%~\cite{Belanger:2013kya}.  
In our model, we have
\begin{eqnarray}
\Gamma\left(h\to X_I X_I\right) 
& = & \frac{\lambda_{HX}^{2}v^{2}}{32\pi M_{h}},
\end{eqnarray}
and constraining ${\rm BR}\left(h\to X_I X_I\right) \lesssim 0.2$, 
we obtain 
\begin{equation}
\lambda_{HX} \lesssim 0.01.
\label{eq:HX_constraint}
\end{equation} 
On the other hand, for a gauged $U(1)_X$, 
there will be no such decay channel since the gauge 
boson associated with $U(1)_X$ breaking is much heavier than the SM Higgs
with its mass given by
\begin{eqnarray}
M_{h}^{2} & = & \lambda_{H}\left(1+\delta_{HX}\right)v^{2}.
\end{eqnarray}

For simplicity, we assume $\delta_{HX} \ll 1$ such that 
the scalars $Y_R$, $Y_I$, $X_R$ and $X_I$ are approximately 
mass eigenstates with respective masses
\begin{equation}
M_{Y_{R}}^{2} = M_{Y_{I}}^{2} = M_{Y}^{2},\;\;\;\;
M_{X_{R}}^{2} = \lambda_{X}w^{2},\;\;\;\;
M_{X_{I}}^{2} = 0\, .
\label{eq:scalar_masses}
\end{equation}

We assume $M_{X_{R}}\gg M_{Y_{R}}$ such that $Y_{R}$ cannot decay to
$X_{R}$ but it can decay directly to $2h$ or 
indirectly to $4h$ through two off-shell $X_{R}$. 
In the following, we will drop the subscript $R$ for the LLP $Y_R$
and simply denote it as $Y$. From the context, there should be no confusion 
with the original complex field $Y$. 

The decay widths for the decays of $Y$ to scalars are given by
\begin{eqnarray}
\Gamma\left(Y \to X_{I}X_{I}\right) & = & \frac{1}{32\pi}\frac{\left(\lambda_{XY}u + \mu_{XY} \right)^{2}}{M_{Y}}\, , 
\label{eq:Decay_width_YXX}\\
\Gamma\left(Y \to 2h\right) & = & \frac{\lambda_{HY}^{2}}{32\pi}\frac{u^{2}}{M_{Y}},
\label{eq:Decay_width_Yhh}\\
\Gamma\left(Y \to 4h\right) & \approx & \frac{\lambda_{HX}^{4}}{16384\pi^{5}}\left(\frac{\lambda_{XY}u - \mu_{XY}}{\lambda_{X}}\right)^{2}\frac{M_{Y}^{3}}{M_{X_{R}}^{4}}\, .
\label{eq:Decay_width_Yhhhh}
\end{eqnarray}
Comparing the decay rates, we have
\begin{eqnarray*}
\frac{\Gamma\left(Y \to 4h\right)}{\Gamma\left(Y \to X_{I}X_{I}\right)}
 & = & \frac{1}{512 \, \pi^{4}} 
\left(\frac{\lambda_{HX}}{\lambda_X}\right)^{4}
\left(\frac{\lambda_{XY} - r_Y^2}{\lambda_{XY} + r_Y^2}\right)^{2}r_Y^{4} \, , \nonumber \\
\frac{\Gamma\left(Y \to 4h\right)} {\Gamma\left(Y \to 2h\right)}
 & = & \frac{1}{512\,\pi^{4}}
\left(\frac{\lambda_{HX}}{\lambda_X}\right)^{4}
\left(\frac{\lambda_{XY} - r_Y^2}{\lambda_{HY}}\right)^{2}r_Y^{4}.\nonumber 
\end{eqnarray*}
where we define $r_Y \equiv M_Y/w$ and we have used 
Eq.~\eqref{eq:scalar_masses} for the mass $M_{X_R}$
and also Eq.~\eqref{eq:Y_vev}.
Taking $10^9\,{\rm GeV} \lesssim w \lesssim 10^{12}\,{\rm GeV}$
and $M_Y \sim 10^6$ GeV, we have $10^{-6} \lesssim r_Y \lesssim 10^{-3}$.
With the assumption $M_{X_R} \gg M_Y$, we have 
$\lambda_X \gg r_Y^2$. 
However with the assumption of small $H-X$ mixing $\delta_{HX} \ll 1$, 
we have $\lambda_{HX}/\lambda_X \ll \lambda_H/(4\lambda_{HX})
\simeq M_h^2/(4\lambda_{HX}v^2)$. Taking the maximum allowed value 
$\lambda_{HX} = 10^{-2}$ from Eq.~\eqref{eq:HX_constraint},
we have $\lambda_{HX}/\lambda_X \ll 13$.
As we will see later in Section \ref{subsec:LLP_DM}, 
under reasonable assumptions, $\lambda_{HY}$ and $\lambda_{XY}$ are 
required to be small, $\lesssim 10^{-10}$, in order not to over-produce $Y$.
Taking all the above considerations into account, we can write 
down conservative upper bounds
\begin{equation*}
\frac{\Gamma\left(Y \to 4h\right)}{\Gamma\left(Y \to X_{I}X_{I}\right)} \lesssim 6\times10^{-13},\;\;\;\;
\frac{\Gamma\left(Y \to 4h\right)}{\Gamma\left(Y \to 2h\right)} \lesssim 6\times10^{-13}\left(\frac{\lambda_{XY}-r_Y^2}{\lambda_{HY}}\right)^{2}\, .
\end{equation*}
From the above, we see that in order to have 
$\Gamma (Y \to 4h) > \Gamma(Y \to 2h)$, we need a coupling 
$\lambda_{HY}\lesssim 7 \times 10^{-7}|\lambda_{XY} - r_Y^2|$. 
In addition, we also notice that the decay channel of $Y$ to the NGB $X_I$ 
always dominates over the channel $Y \to 4h$ while it is generally 
faster than $Y \to 2h$ unless $\lambda_{HY} > |\lambda_{XY} + r_Y^2|$.
Hence in the global $U(1)_X$ scenario, this channel $Y \to X_I X_I$ 
is usually the one which determines the lifetime of $Y$.\footnote{In this
case, since $Y$ has additional decay channel to invisible NGBs, the LLP 
lifetime $\tau_Y$ in the fit of the neutrino flux in the previous sections 
will be the partial lifetime of $Y$.}
Requiring $\tau_{Y} > t_{0}\simeq 4.4\times10^{17}\,{\rm s}$, 
we obtain using Eq.~\eqref{eq:Decay_width_YXX}
\begin{eqnarray}
\mu_{XY} & \lesssim & 
1.2\times10^{-17} \left( 1 + \frac{\lambda_{XY}}{r_Y^2}\right)^{-1}
\left(\frac{M_Y}{10^6\,{\rm GeV}}\right)^{1/2}{\rm GeV}.
\label{eq:muXY_bound}
\end{eqnarray}
The constraint above does not apply in the gauged $U(1)_X$ scenario
since the $U(1)_X$ gauge boson is assumed to be a lot heavier than $Y$.
From Section \ref{sec:fitting}, we see that the lifetimes
which fit the data fall in the range $\tau_Y \sim 10^{27-29}$ s, 
assuring $Y$ to be long-lived. In this case, we can constrain
$\mu_{XY}$ using Eq.~\eqref{eq:Decay_width_Yhh}
and obtain
\begin{eqnarray}
\mu_{XY} & \lesssim & 
2.6\times10^{-19} \left(\frac{M_Y}{10^6\,{\rm GeV}}\right)^{5/2}
\left(\frac{10^{10}\,{\rm GeV}}{w}\right)^{2}
\left(\frac{10^{-11}}{\lambda_{HY}}\right)
{\rm GeV}.
\label{eq:muXY_bound2}
\end{eqnarray}
The fact that the bound \eqref{eq:muXY_bound2} is smaller than the
bound \eqref{eq:muXY_bound} and not the other way around is
interesting and also \emph{crucial}.  It implies that in the global
$U(1)_X$ scenario, if we can fit the neutrino flux to explain the
IceCube excess, despite having a dominant decay channel of $Y$ to NGB,
the longevity requirement on $Y$ is automatically fulfilled.  In all
cases, we see that $\mu_{XY}$ is constrained to be very small and the
condition in obtaining the induced vev \eqref{eq:Y_vev} is always
valid.

\subsection{The Fermionic Sector}

Now we describe the new fermionic sector of our model. 
With the introduction of a pair of vectorlike fermion doublets $\Psi_L$,
$\Psi_R$ and a right-handed singlet $N_R$, we have the following new terms
\begin{eqnarray}
-{\cal L} & \supset & \left(y_{\Psi}\overline{\ell_{L}}\Psi_{R}Y
+y_{\nu}\overline{\Psi_{L}}\widetilde{H}N_{R}
+M_{\Psi}\overline{\Psi_{L}}\Psi_{R}+{\rm H.c.}\right)\, ,
\label{eq:fermion_lag}
\end{eqnarray}
where $\ell_L = (\nu_L,e_L)^T$ are the SM lepton doublets with the
flavour index suppressed and $\widetilde{H}=i \sigma_2 H^{*}$ with
$\sigma_2$ the Pauli matrix.  We assume that $M_\Psi > $ PeV such that
$Y$ cannot decay into it.  After EW symmetry breaking, a
mixing of the new fermions with the SM leptons will be induced such
that
\begin{eqnarray}
{\cal L}_{m} & = & 
\left(\begin{array}{cc}
\overline{e_{L}} & \overline{\psi_{L}^{-}}\end{array}\right)m_{e\Psi}\left(\begin{array}{c}
e_{R}\\
\psi_{R}^{-}
\end{array}\right) +
\left(\begin{array}{cc}
\overline{\nu_{L}} & \overline{\psi_{L}^{0}}\end{array}\right)m_{\nu\Psi}\left(\begin{array}{c}
N_{R}\\
\psi_{R}^{0}
\end{array}\right)+{\rm H.c.}\, ,
\end{eqnarray}
where we have defined the mass matrices
\begin{eqnarray}
m_{e\Psi} & = & \left(\begin{array}{cc}
y_{e}v & y_{\Psi}u\\
0_{1 \times 3} & M_{\Psi}
\end{array}\right)\, , \\
m_{\nu\Psi} & = & \left(\begin{array}{cc}
0_{3\times 1} & y_{\Psi}u\\
y_{\nu}v & M_{\Psi}
\end{array}\right)\, ,
\end{eqnarray}
with $y_e$ the $3\times 3$ SM charged lepton Yukawa matrix and 
$y_{\Psi}$ a 3-column vector. Without loss of generality, 
we can choose a basis where $y_e = \hat y_e$ is diagonal and real. 
Diagonalizing the mass matrix for charged leptons above, 
due to the very small $u/M_{\Psi}$, the charged leptons mass eigenstates 
are still $m_\alpha \equiv (\hat y_e)_{\alpha\alpha}v$ to a good approximation.
For the neutral leptons, since we have introduced only one $N_R$, 
there is only one massive active neutrino with Dirac mass given by
\begin{eqnarray}
m_{\nu} & = & \sqrt{\sum_{\alpha}\left|\left(y_{\Psi}\right)_{\alpha}\right|^{2}}\,
u\frac{y_{\nu}v}{M_{\Psi}}\, ,
\label{eq:nu_mass}
\end{eqnarray}
As we will see shortly, the longevity of $Y$ implies an extremely
small contribution to neutrino mass.  This could be easily modified to
accommodate the neutrino oscillation data for example by introducing
two other heavy right-handed SM singlets uncharged under $U(1)_X$ or
more generally by introducing a dimension five Weinberg
operator~\cite{Weinberg:1979sa}.  Since this is not directly relevant
for our current study, we won't pursue it further.

Next the leptonic decay widths of $Y$ to charged and neutral leptons
are, respectively,
\begin{eqnarray}
\Gamma\left(Y \to e_{{\alpha}}\overline{e_{{\beta}}}\right) 
& = & \frac{1}{32\pi}\left|\left(y_{\Psi}\right)_{\alpha}\right|^{2}
\left|\left(y_{\Psi}\right)_{\beta}\right|^{2}
\frac{u^{2}{m}_{\beta}^{2}}{M_{\Psi}^{4}}M_{Y}\, ,
\end{eqnarray}
and 
\begin{eqnarray}
\Gamma\left(Y \to\nu_{{\alpha}}\overline{N_{R}}\right) 
& = & \frac{1}{32\pi}\left|\left(y_{\Psi}\right)_{\alpha}\right|^{2}
\frac{\left|y_{\nu}\right|^{2}v^{2}}{M_{\Psi}^{2}}M_{Y}\, .
\end{eqnarray}
Taking the ratio of these widths we have
\begin{eqnarray*}
\frac{\Gamma\left(Y \to e_{{\alpha}}\overline{e_{{\beta}}}\right)}
{\Gamma\left(Y\to\nu_{{\delta}}\overline{N_{R}}\right)} 
& = & \left|\frac{\left(y_{\Psi}\right)_{\alpha}}{\left(y_{\Psi}\right)_{\delta}}\right|^2
\left|\frac{\left(y_{\Psi}\right)_{\beta}}{y_{\nu}}\right|^2 \left(\frac{u}{M_{\Psi}}\right)^2 
\left(\frac{{m}_{\alpha}}{v}\right)^2 \, .
\end{eqnarray*}
Assuming the dimensionless couplings are of the same
order and considering, for example, $m_\tau/v$, 
the ratio above is still suppressed by very small $u/M_{\Psi}$.
Hence the decays of $Y$ into neutrinos will 
always largely dominate over the decays into charged leptons 
and the latter can be ignored.

Using Eq.~\eqref{eq:nu_mass}, the total decay width of 
$Y$ to neutrinos can be rewritten as 
\begin{eqnarray*}
\sum_\alpha \Gamma\left(Y \to\nu_{{\alpha}}\overline{N_{R}}\right) 
& = & \frac{1}{32\pi}\frac{m_{\nu}^{2}}{u^{2}}M_{Y}\,.
\end{eqnarray*}
Requiring the lifetime of $Y$ to be longer 
than the age of the Universe $t_{0}$, we have 
\begin{equation}
\frac{m_{\nu}}{u} = \frac{m_{\nu} r_Y^2}{\mu_{XY}} \lesssim 10^{-23}.  
\label{eq:bound_mnu}
\end{equation}
In order to maximize the contribution to neutrino mass,  
we take for instance $r_Y = 10^{-6}$ and $\mu_{XY} = 10^{-17}$ GeV
(see Eqs.~\eqref{eq:muXY_bound} and \eqref{eq:muXY_bound2}),
we still only have $m_{\nu} \lesssim 10^{-19}$ eV. 
We can also translate the bound \eqref{eq:bound_mnu} into a bound 
for the dimensionless couplings using Eq.~\eqref{eq:nu_mass} as follows
\begin{equation}
\sqrt{\sum_{\alpha}\left|\left(y_{\Psi}\right)_{\alpha}\right|^{2}}\,y_{\nu}
\lesssim 6 \times 10^{-20} \left( \frac{M_{\Psi}}{10^6\; {\rm GeV}}\right).
\label{eq:bound_y}
\end{equation}
As we will see in Section \ref{subsec:LLP_DM}, for $M_{\Psi} \sim$ PeV, 
we need $(y_{\Psi})_\alpha \lesssim 10^{-11}$ in order not to over-produce $Y$
and this in turn implies $y_{\nu} \lesssim 10^{-9}$.

Finally, we would like to comment on possible 
phenomenological constraints on the fermion sector.
In Ref.~\cite{Ishiwata:2013gma}, it was shown that 
the contributions to electric dipole moment of the electron 
and charged lepton flavour violating processes are much suppressed
beyond the current bounds if new heavy vectorlike leptons have masses
beyond $\sim 100$ TeV. In our scenario, besides having $M_{\Psi} \gg 100$ TeV,
the constraints on the model parameters from the requirement of 
the longevity of $Y$ render these phenomenological 
constraints completely irrelevant.

\subsection{Does the Long-lived Particle Constitute Most of the DM?}
\label{subsec:LLP_DM}

Here we would like to study the abundance of the LLP in our model.
The abundance of the LLP $Y$ is bounded from above by the DM abundance
of the Universe. Clearly the abundance of a PeV LLP depends on the
re-heating temperature $T_{RH}$ of the Universe after inflation.  If
$T_{RH} \ll M_Y$, no $Y$ would be generated.  If $T_{RH} \sim M_Y$,
some amount of $Y$ would be generated.  If $T_{RH} \gg M_Y$, it is
possible to generate significant amount of $Y$.  Here we will 
consider the last and most constraining scenario and determine the
constraints on our model parameters in order not to over-produce $Y$.
First let us estimate what is the upper bound on a PeV $Y$
abundance. From Planck 2013 measurement, the ratio between DM 
over baryon densities are~\cite{Ade:2013zuv}
\begin{equation}
\frac{\Omega_{DM}}{\Omega_{B}} = \frac{Y_{DM} M_{DM}}{Y_B M_B} = 5.3,
\end{equation}
where we denote $Y_{x} \equiv {n_x/s}$, the $x$ number density
normalized by the entropy density $s = (2\pi^2/45)g_\star T^3$.
During the radiation dominated period of the Universe, the relativistic
degrees of freedom $g_\star = 106.75$, assuming only the SM particles.
Taking the baryon mass $M_B = 1$ GeV and $Y_B = 8.8\times10^{-11}$, we
have
\begin{equation}
Y_{DM} = 1.2\times 10^{-16 }\left(\frac{4\times 10^6\,{\rm GeV}}{M_{DM}}\right).
\label{eq:PeV_DM_abundance}
\end{equation}
Hence the above is the upper bound for the abundance $Y_Y$ of our
LLP $Y$ of mass $M_Y$.

\begin{figure}[htb]
\begin{center}
\includegraphics[scale=0.53]{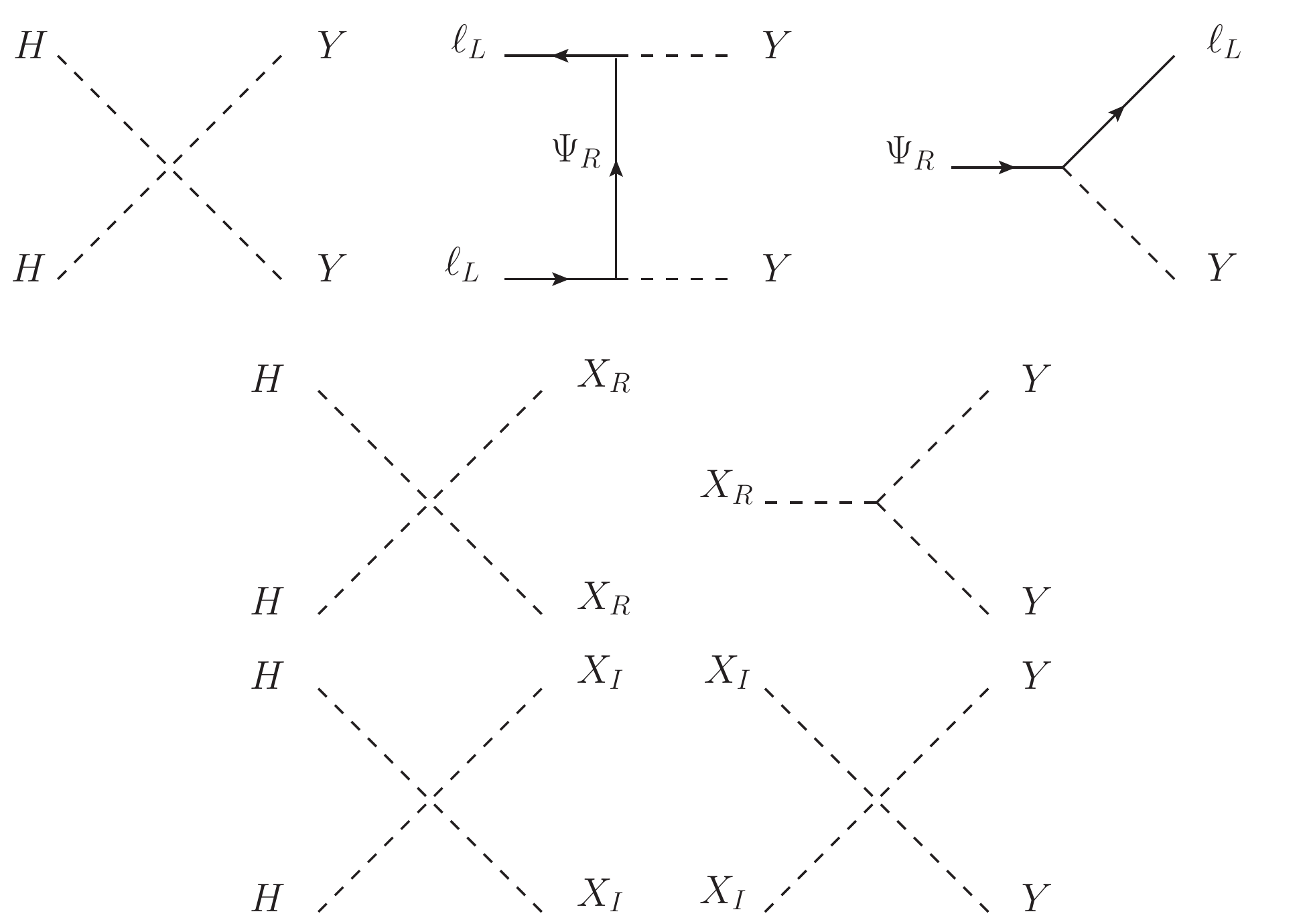}
\vspace{-2mm}
\end{center}
\vspace{-4mm}
\caption{The scattering and decay processes which help to populate $Y$
  abundance in the early Universe. For gauged $U(1)_X$, only the
  processes in the top two rows are relevant while for global
  $U(1)_X$, we also have to include the scattering processes with NGB
  in the last row.  For the t-channel scattering processes, it is
  understood that we also have to include the u-channel scattering
  processes.}
\label{fig:scatterings}
\end{figure}

First, let us discuss the possibility to have the correct $Y_Y$ in the
thermal freeze-out scenario.  In this scenario, the scattering
processes $Y Y \leftrightarrow H H$, $Y Y \leftrightarrow \ell_L
\ell_L$ and $Y Y \leftrightarrow X_I X_I$ (the last scattering process
is relevant only for global $U(1)_X$ scenario) as shown in
Figure~\ref{fig:scatterings} are fast to keep $Y$ in thermal
equilibrium at $T \gg M_Y$. Since the EW symmetry remains unbroken, 
$\ell_L$ and $H$ refer to the SM lepton and Higgs
doublets respectively.  For gauged $U(1)_X$, if $T_{RH} \gg M_g$ with
$M_g$ the $U(1)_X$ gauge boson mass, this gauge interactions can also
keep $Y$ in thermal equilibrium. The final $Y_Y$ can be estimated from
the time when these scattering processes freeze out. In the following
we will define the temperature as $z \equiv M_Y/T$. Due to fast
reactions which keep $Y$ in equilibrium, $Y_Y$ is equal to its
equilibrium abundance $Y_Y^{eq}(z) = 45/(2\pi^4 g_\star)z^2 {\cal
  K}_2(z)$ with ${\cal K}_2(z)$ the modified Bessel function of type 2
until the time of freeze-out at $z_{fo} \equiv M_Y/T_{fo}$. Hence we
can estimate the final $Y_Y \approx Y_Y^{eq}(z_{fo}) \approx 10^{-16}$
which gives us $z_{fo} \approx 35$.  Such a late freeze-out implies
large annihilating cross sections.  We found that for $M_Y \sim$ PeV,
such large cross sections cannot be obtained if $\lambda_{HY}$ and
$y_\Psi$ (also $\lambda_{XY}$ for global $U(1)_X$ case) are bounded to
be perturbative i.e. $< 4\pi$.  Since thermal freeze-out is not a
viable scenario, for gauged $U(1)_X$, we have to ensure that the gauge
interactions between $Y$ and $U(1)_X$ gauge boson are not in thermal
equilibrium.  Since the gauge coupling is generically large, we can
only suppress this interaction by assuming $T_{RH}$ much smaller than
the mass of the gauge boson $M_g$ such that there is no gauge boson
for gauge scatterings to take place.

In the following, we will study the scenario in which
$Y$ abundance is obtained from the freeze-in scenario 
where $Y$ is very weakly coupled to other particles in the thermal bath
and is generated from the interactions with those particles~\cite{Hall:2009bx}. 
With the assumption $T_{RH} \ll M_g$, the processes
we have to consider are shown in Figure~\ref{fig:scatterings}. 
We only show t-channel scattering processes
and it is understood that in the calculation, one should also include 
the u-channel scattering processes. For gauged $U(1)_X$, we have
to consider processes in the top two rows of Figure~\ref{fig:scatterings}: 
$H H \leftrightarrow Y Y$, $\bar\ell_L \ell_L \leftrightarrow Y Y$, 
$\Psi_R \leftrightarrow \ell_L Y$, $H H \leftrightarrow X_R X_R$ 
and $X_R \leftrightarrow Y Y$. For global $U(1)_X$, 
we also have to take into account the scattering with the NGB 
and consider two additional scatterings 
$H H \leftrightarrow X_I X_I$ and $X_I X_I \leftrightarrow Y Y$.

The SM particles $H$ and $\ell_L$ are necessarily in thermal
equilibrium due to fast gauge interactions. 
If $T_{RH} \gg M_\Psi$, the vectorlike fermion doublets $\Psi_L$, $\Psi_R$ 
which are charged under $SU(2)_L \times U(1)_Y$ will also be 
in thermal equilibrium due to fast gauge interactions.
If $T_{RH} \gg M_{X_R}$, taking the value allowed by invisible
Higgs decay constraint \eqref{eq:HX_constraint} $\lambda_{HX} = 10^{-2}$,
$X_R$ will also be thermal equilibrium due to the fast scatterings
$H H \leftrightarrow X_R X_R$. Finally for the case of 
global $U(1)_X$, taking $\lambda_{HX} = 10^{-2}$, 
$X_I$ will also be in thermal equilibrium.
On the other hand, $Y$ will be populated due to the scatterings 
$H$, $\ell_L$ and $X_I$ and the decays of $\Psi_R$ and $X_R$
as shown in Figure~\ref{fig:scatterings}. We can write down 
the Boltzmann equation to describe the generation of $Y$
\begin{eqnarray}
s{\cal H}z\frac{dY_Y}{dz} & = & 2\left(\gamma_{HH\leftrightarrow YY}
+\gamma_{\ell_L\ell_L\leftrightarrow Y Y}\right)
\left[1-\left(\frac{Y_Y}{Y_Y^{eq}}\right)^2\right] 
+ \gamma_{\Psi_R \leftrightarrow \ell_L Y}\left[1-\frac{Y_Y}{Y_Y^{eq}}\right] \nonumber \\
& & + 2\left(\gamma_{X_R\leftrightarrow YY}+ \gamma_{X_I X_I \leftrightarrow YY}\right)
\left[1-\left(\frac{Y_Y}{Y_Y^{eq}}\right)^2\right],
\label{eq:BE}
\end{eqnarray}
where $\gamma_{a}$ is the thermal averaged reaction density for the
corresponding process $a$ and ${\cal H} =
\sqrt{4\pi^3/45}\sqrt{g_\star}T^2/M_{\rm Pl}$ is the Hubble expansion
rate with the Planck mass $M_{\rm Pl}=1.22 \times 10^{19}$ GeV.  In
writing the equation above, we have also taken $H$, $\ell_L$,
$\Psi_R$, $X_R$ and $X_I$ to be in thermal equilibrium.  In fact, since
$y_{\Psi} \ll 1$ (weakly coupled) and
$\gamma_{\ell_L\ell_L\leftrightarrow Y Y} \propto y_{\Psi}^4$ while
$\gamma_{\Psi_R \leftrightarrow \ell_L Y} \propto y_{\Psi}^2$, we will
ignore the sub-dominant scattering process
$\ell_L\ell_L\leftrightarrow Y Y$.  In addition with $Y_Y \ll
Y_Y^{eq}$, we can further simplify Eq.~\eqref{eq:BE} by dropping the
terms with $Y_Y/Y_Y^{eq}$.  Finally we obtain a very simple Boltzmann
equation
\begin{equation}
s{\cal H}z\frac{dY_Y}{dz} = 2\gamma_{HH \leftrightarrow YY}
+\gamma_{\Psi_R \leftrightarrow \ell_L Y} 
+2\gamma_{X_R \leftrightarrow YY} + 2\gamma_{X_I X_I \leftrightarrow YY}.
\label{eq:BE_sim}
\end{equation}
In the above the thermal averaged reaction densities are given by
\begin{eqnarray}
\gamma_{HH \leftrightarrow YY} & = & \frac{\lambda_{HY}^2}{256\pi^5 }\,
M_Y^4 \frac{[{\cal K}_1(z)]^2}{z^2}, \\
\gamma_{\Psi_R \leftrightarrow \ell_L Y} & = & 
\frac{\left|y_{\Psi}\right|^2}{32\pi^3} a_{\Psi}^3 M_Y^4\,
\frac{{\cal K}_1(a_{\Psi} z)}{z} 
\left(1-\frac{1}{a_{\Psi}^2}\right)^2, \\
\gamma_{X_R \leftrightarrow Y Y} & = & \frac{\lambda_{XY}^2}{64\pi^3} a_{X_R} w^2 M_Y^2\,
\frac{{\cal K}_1(a_{X_R} z)}{z} \sqrt{1-\frac{4}{a_{X_R}^2}}, \\
\gamma_{X_IX_I \leftrightarrow YY} & = & \frac{\lambda_{XY}^2}{2048\pi^5 }\,
M_Y^4 \frac{[{\cal K}_1(z)]^2}{z^2},
\end{eqnarray}
where $a_{i} \equiv M_i/M_Y$ and we have summed over lepton flavour
$|y_\Psi|^2 \equiv \sum_\alpha |(y_\Psi)_\alpha|^2$.  We can
analytically solve Eq.~\eqref{eq:BE_sim} and obtain the final $Y$
abundance as follows
\begin{eqnarray}
Y_Y & \simeq & 10^{-16} \left(\frac{106.75}{g_\star}\right)^{3/2} 
\left(\frac{4\times 10^6\,{\rm GeV}}{M_Y}\right)
\left [
\left(\frac{\lambda_{HY}}{4\times 10^{-11}}\right)^2
+ \left(\frac{|y_\Psi|}{8\times 10^{-12}}\right)^2
\left(\frac{10}{a_{\Psi}}\right)
\right. \nonumber \\
& & \!\!\!\!\!\!\!\!\! 
\left. + \left(\frac{\lambda_{XY}}{1\times 10^{-10}}\right)^2
\left(\frac{10^3}{a_{X_R}}\right)^3 
\left(\frac{4\times 10^6\,{\rm GeV}}{M_Y}\right)^2
\left(\frac{w^2}{10^{10}\,{\rm GeV}}\right)^2 
+ \left(\frac{\lambda_{XY}}{1\times 10^{-10}}\right)^2 
\right].
\label{eq:Y_sol}
\end{eqnarray}
In the solution above we have ignored the phase space factors in the
decays which are negligible when $a_{\Psi}, a_{X_R} \gg 1$.  In
Eq.~\eqref{eq:Y_sol}, the contributions in the square bracket are in
the following order $H H \leftrightarrow Y Y$, $\Psi_R \leftrightarrow
\ell_L Y$, $X_R \leftrightarrow Y Y$ and $X_I X_I \leftrightarrow Y
Y$.  The last contribution is only relevant for the global $U(1)_X$
scenario.  We have verified the solution above by numerically solving
Eq.~\eqref{eq:BE}.  All in all, we require $|y_\Psi|, \lambda_{XY},
\lambda_{HY} \lesssim 10^{-11}-10^{-10}$ in order not to have $Y$
exceeding the PeV DM abundance $Y_{DM} \sim 10^{-16}$ as in
Eq.~\eqref{eq:PeV_DM_abundance}.

Finally we would like to reiterate that in the fit of Sections
\ref{sec:fitting} and \ref{sec:future}, we have assumed the LLP
constitutes all the DM. If the LLP constitutes a fraction $\kappa$ of
the DM, then its lifetime has to be shorter by a factor of $\kappa$ in
order to maintain the observed flux by IceCube.  Hence, we cannot
choose an arbitrarily small $\kappa$ and at some point when $\kappa
\tau_Y < t_0$, it will cease to be the LLP.  Since $\tau_Y \propto
\mu_{XY}^{-2}$ (see
Eqs.~\eqref{eq:Decay_width_YXX}--\eqref{eq:Decay_width_Yhhhh}), to
obtain $\kappa \tau_Y$, $\mu_{XY}$ has to increase by factor of
$1/\sqrt{\kappa}$.  In the gauged $U(1)_X$ scenario, we are allowed to
take $\kappa$ as low as $\sim 10^{-10}$. On the other hand, for the
global $U(1)_X$ scenario, due to the fast decay of $Y$ into NGB, we
require $\kappa \gtrsim 10^{-4}$ which can be inferred from
Eqs.~\eqref{eq:muXY_bound} and \eqref{eq:muXY_bound2}.  In other
words, the LLP has to constitute at least about 0.01 \% of the DM in
the global $U(1)_X$ scenario.

\section{Conclusion}
\label{sec:conc}
The serendipitous discovery of high-energy extraterrestrial  
neutrinos by IceCube inaugurates an extremely exciting era for 
neutrino astronomy. It is clear that the neutrino candidate events  
above 60 TeV cannot be explained by atmospheric
neutrinos alone. Moreover, the absence of events in the region 500 TeV - 1 PeV 
and above 2 PeV seem to suggest a neutrino spectrum beyond a 
single power-law. This will have to be confirmed or rejected by future data.

We investigate the possibility that the IceCube neutrino events with
energies from 30 TeV to 2 PeV can be explained by (a) solely a
power-law spectrum, (b) decays of a PeV scale LLP with power-law
spectrum component and (c) solely decays of a PeV scale LLP.  For
scenario (c), we study a simple scenario where a scalar LLP, $Y$,
decays dominantly into $Y\to \nu N$, $Y\to 2 h$ or $Y\to 4h$.  We
present a simple extension of the SM where two extra complex scalars,
singlets under the SM group, a vectorlike pair of fermion doublets and
a right-handed fermion singlet are charged under a global or gauged
$U(1)_X$. We show that this model can give rise to a LLP that can
decay dominantly to the above modes while the LLP abundance
is generated through the freeze-in mechanism. In particular, we find that
if $U(1)_X$ is a global symmetry, such LLP has to constitute at least
0.01\% of the total DM of the Universe due to the long-lived
constraint.

Using the current IceCube three-year data set of 37
events~\cite{Aartsen:2014gkd}, we find that all the three scenarios
(a)--(c) above fit the data equally well due to the low statistics.
Our results for these fits are summarized in
Table~\ref{tab:Summary-of-Pvalues}.  In order to disentangle various
scenarios above, we simulate the future IceCube data based on current
hypotheses up to neutrino energies of 10 PeV. 
Using only the energy spectrum information, we determine the exposure time 
needed to disentangle various scenarios by future data (summarized in
Table~\ref{tab:Exclusion-Time.}).  Assuming a power-law spectrum, the
solely LLP decay models can be excluded with two to five times the current
data. Other combinations will take longer time and in some cases 
it will be impossible to distinguish between hypotheses.

\begin{acknowledgments}
This work was supported by Funda\c{c}\~ao de Amparo \`a Pesquisa do
Estado de S\~ao Paulo (FAPESP) and Conselho Nacional de Ci\^encia e
Tecnologia (CNPq).  H.M. thanks Universidade de S\~ao Paulo for the
great opportunity of stay under ``Programa de Bolsas para Professors
Visitantes Internacionais na USP''.  H.M. and R.Z.F. thank the Kavli
Institute for Theoretical Physics in UC Santa Barbara for its
hospitality, where part of this work was completed: This research was
supported in part by the National Science Foundation under Grant
No. NSF PHY11-25915. We also acknowledge partial support from the
European Union FP7 ITN INVISIBLES (Marie Curie Actions,
PITN-GA-2011-289442). 
We are grateful to Shigeru Yoshida for informative correspondences,
and H.M. and R.Z.F. thank Kohta Murase for discussions. 
C.S.F. thanks Jordi Salvado for helpful discussions.

\end{acknowledgments}

\appendix
\section{Some Details on the Statistical Treatment}
\label{details}
For the statistical treatment we mainly follow \cite{Lyons:1986em,Agashe:2014kda}.
Through this work we consider several hypotheses in order to explain
the data observed by IceCube. In this appendix we refer to any of these
hypotheses as ${\rm H_0}(\theta_{i})$, where $\theta_{i}$ are the free
parameters of the hypotheses. In order to estimate the preferred
values of $\theta_{i}$, we maximize the likelihood function

\begin{equation}
L(\theta_{i};k_{n})=\prod_{n=1}^{12}f(k_{n};\mu_{n}(\theta_{i}))\, ,
\end{equation}
\noindent with respect to $\theta_{i}$, where
$f(k_{n};\mu_{n}(\theta_{i}))$ is the Poisson probability distribution
function for measuring $k_{n}$ events in the bin $n$ assuming that the
mean value is given by $\mu_{n}(\theta_{i})$.  Notice that the results
obtained from this procedure are equivalent to those obtained from the
minimization of the function
$\chi^{2}(\theta_{i};k_{n})=-2\ln(L(\theta_{i};k_{n}))$.

Therefore, for simplicity and numerical stability, we normally use
$\chi^{2}(\theta_{i};k_{n})$ to report our results. The estimators of
the parameters $\theta_{i}$ are defined as $\hat{\theta}_{i}$, the
minimum of the statistic is given by
$\chi_{\rm min}^{2}\equiv\chi^{2}(\hat{\theta}_{i},k_{n})$ and the mean
values of the hypothesis ${\rm H_0}(\hat{\theta}_{i})$ are given by
$\mu_{n}(\hat{\theta}_{i})$. In the following we show how we
compute the p-value associated to the observed $\chi_{\rm min}^{2}$ and
the corresponding confidence intervals for the variables $\theta_{i}$.

\subsection{p-value}

In order to quantify the level of agreement or incompatibility of
the hypothesis ${\rm H_0}(\hat{\theta}_{i})$ with respect to the current
data, it is useful to evaluate the probability of obtaining the current
value of $\chi_{\rm min}^{2}$ assuming that the data is indeed generated
by the hypothesis ${\rm H_0}(\hat{\theta}_{i}).$

In practice, we must construct the probability distribution function,
$f(t|{\rm H_0})$, with $t=\chi_{\rm min}^{2}(\theta_{i},\bar{k}_{n})$ and
\textbf{$\bar{k}_{n}\sim\mbox{\ensuremath{Pois}}(\mu_{n}(\hat{\theta}))$}.
In general, this function is given by the normalized frequency
distribution of $t$ obtained from several random realizations of
${\rm H_0}(\hat{\theta})$.  Using $f(t|{\rm H_0})$ we are able to compute the
p-value associated to $\chi_{\rm min}^{2}$, which is defined as the
probability to find $t$ in the region of lesser or equal
incompatibility with ${\rm H_0}$ than the level of incompatibility
observed with the current data,

\begin{equation}
p=\int_{\chi_{\rm min}^{2}}^{\infty}f(t|{\rm H_0})dt\, .
\end{equation}

Thus, a very small value of $p$ implies that the observed level of
incompatibility is quite unlikely to be found, which also suggest
that ${\rm H_0}$ is not a good representation of the data. When this
is the case, it is said that the hypothesis ${\rm H_0}$ is rejected at
$(1-p)$ confidence level.

\subsection{Intervals}

For simplicity, we compute the intervals using a Bayesian point of
view. From Bayes theorem, the probability density function of the
parameters $\theta_{i}$ can be obtained from the product of the likelihood
function $L(\theta_{i};k_{n})$ and the joint prior of the parameters
$\theta_{i}$, which we define as $\pi(\theta_{i})$. We assume that
each prior is constant in a finite interval and null otherwise. Thus,
the domain of the variables $\theta_{i}$ is given by a finite hyper
volume $U$, which is obtained from the product of the priors. The
normalized p.d.f. for $\theta_{i}\in U$ is given by,

\begin{equation}
p(\theta_{i})=\frac{L(\theta_{i};k_{n})}{\int_{U}L(\bar{\theta}_{i};k_{n})d\bar{\theta_{i}}}\, .
\label{eq:pdf-interval}
\end{equation}

From the above expression, it is direct to compute the coverage probability
of a region of parameters $V\in U$, since we just have to integrate
the previous expression on the region $V.$ However, it is not direct
to obtain an interval such that the coverage probability is some fixed
number $(1-\alpha)$, because this procedure involves an integral
equation. Furthermore, it is not easy to find a rule to discriminate
within degenerate solutions. These problems can be addressed in a
simple way when we consider the discrete form of Eq. (\ref{eq:pdf-interval}).

Therefore, we generate a quite big sample of points $\theta_{i}$
covering the domain $U$. In general we use
$N=(10^{4},500^{2},200^{3})$ when the number of free parameters is
$n=(1,2,3)$ respectively. In the same procedure we are able to compute
the pairs $\{\theta_{i},L(\theta_{i};k_{n})\}_{j}$, with
$j=1..N$. Using the elements of this set we can compute the coverage
probability of some particular region, but first we need to fix the
rule to discriminate within degenerate intervals. For instance, we
require that the resulting interval contains the most likely points.
This is analogous to considering a symmetric interval around the center of
a Gaussian distributed variable. To implement this requirement, we
just order the set $\{\theta_{i},L(\theta_{i};k_{n})\}_{j}$ in
descending order in $L(\theta_{i};k_{n})$ such that the new set satisfies
the condition $L(\theta_{i};k_{n})_{j}\geq L(\theta_{i};k_{n})_{k}$
with $k=j+1..N$. Finally, the hyper volume with $(1-\alpha)$ coverage
probability is given by the set
$\{\theta_{i},L(\theta_{i};k_{n})\}_{j}$ with $j=1..M,$ such that

\begin{equation}
\label{eq:coverage}
\frac{\sum_{j=1}^{M}L(\theta_{i};k_{n}){}_{j}}{\sum_{l=1}^{N}L(\theta_{i};k_{n}){}_{l}}\leq(1-\alpha) \, ,
\end{equation}
\noindent where we have assumed that the volume element $\Delta\theta_{i}$
is constant everywhere. The final procedure only involves a loop on
the variable $j$ that runs in unit steps until the condition \eqref{eq:coverage}
is violated, which determines the value of $M$. The intervals at
some confidence level for a given variable $\theta_{i}$ correspond
to the extreme values of the set $\{\theta_{i},L(\theta_{i};k_{n})\}_{j}$
with $j=1..M$. 

\subsection{Exclusion Time}

In this section we compute the necessary exposure time in order to
exclude a given hypothesis ${\rm H_j}$ assuming that the hypothesis ${\rm H_i}$
is the true explanation of the current data. This procedure is based
in the computation of p-values between pairs of hypotheses at different
times. Below, we describe the algorithm step by step: 
\begin{enumerate}
\item In general, we compute the mean values of some hypothesis
  ${\rm H_0}(\theta_{i})$ at a time $T=f_{T}\times 988$ days from the
  product of the mean values predicted today, defined as
  $\mu_{n}(\theta_{i})$, times the factor $f_{T}$. Thus,
  $\mu_{n}(\hat{\theta}_{i},T)=\mu_{n}(\hat{\theta}_{i})\times f_{T}$.
  For simplicity we call the version of ${\rm H_0}(\theta_{i})$ at a time
  $T$ as ${\rm H_0}(\theta_{i},T)$.
\item Assuming that the hypothesis ${\rm H_i}(\hat{\theta_{i}},T)$ is the true
hypothesis, we compute the number of expected pseudo events $\bar{k}_{n}(T)$
at a fixed time $T$. As these numbers follow a random distribution,
such that $\bar{k}_{n}(T)\sim Pois(\mu_{n}(\hat{\theta}_{i},T))$,
we notice that there is not a unique way to determine their values.
Then, in order to define a stable algorithm we just pick the most
likely values of $\bar{k}_{n}(T)$ given $\mu_{n}(\hat{\theta_{i}},T)$.
This approach covers the most likely outputs of an hypothetical experiment
and it is stable under repetitions of the procedure.
\item We take the hypothesis ${\rm H_j}(\theta_{i},T)$ with $f_{T}$ fixed
  and $\theta_{i}$ free. We estimate the best fit parameters of the
  hypothesis ${\rm H_j}(\theta_{i},T)$ using the likelihood method with
  respect to the observed pseudo events generated from
  ${\rm H_i}(\hat{\theta_{i}},T)$.  Finally, we compute the p-value of
  this fit. We choose a minimum p-value such that both hypotheses are
  incompatible. Following a conservative approach we consider $p=0.05$
  as the threshold.
\end{enumerate}

\newpage

\section{Confidence Intervals}
\label{summary}
The summary of the confidence intervals for each hypothesis $\rm H_0$
are showed in Table \ref{tab:Confidence-level-intervals}.
\begin{table}[hb!]
\begin{centering}
\begin{tabular}{|c|c|c|c|c|c|}
\hline 
${\rm H_0}$ & Best fit & $U$ (prior) & $\pm1\sigma$ $(68\%)$  & $\pm2\sigma$ $(95\%)$ & $\pm3\sigma$ $(99\%)$\tabularnewline
\hline 
\hline 
\multirow{4}{*}{I} & \multirow{2}{*}{$s=2.3$} & \multirow{2}{*}{$[0,4]$} & $2.83$ & $3.21$ & $3.46$\tabularnewline
\cline{4-6} 
 &  &  & $1.72$ & $1.12$ & $0.57$\tabularnewline
\cline{2-6} 
 & \multirow{2}{*}{$C_{0}=0.6$} & \multirow{2}{*}{$[10^{-4},10]^{*}$ } & $1.41$ & $2.08$ & $2.51$\tabularnewline
\cline{4-6} 
 &  &  & $0.12$ & $0.01$ & $1.33\times10^{-3}$\tabularnewline
\hline 
\multicolumn{6}{|c|}{}\tabularnewline
\hline 
\multirow{6}{*}{II.a} & \multirow{2}{*}{$s=2.43$} & \multirow{2}{*}{$[0,4]$ } & $3.5$ & $4.0$ & $4.0$\tabularnewline
\cline{4-6} 
 &  &  & $0.38$ & $0.0$ & $0.0$\tabularnewline
\cline{2-6} 
 & \multirow{2}{*}{$C_{0}=0.51$} & \multirow{2}{*}{$[10^{-4},10]^{*}$ } & $1.66$ & $2.1$ & $2.35$\tabularnewline
\cline{4-6} 
 &  &  & $2.25\times10^{-3}$ & $10^{-4}$ & $10^{-4}$\tabularnewline
\cline{2-6} 
 & \multirow{2}{*}{$\tau_Y=5.26$} & \multirow{2}{*}{$[0.01,100]^{*}$ } & $100$ & $100$ & $100$\tabularnewline
\cline{4-6} 
 &  &  & $1.17$ & $0.85$ & $0.74$\tabularnewline
\hline 
\multicolumn{6}{|c|}{}\tabularnewline
\hline 
\multirow{6}{*}{II.b} & \multirow{2}{*}{$s=2.76$} & \multirow{2}{*}{$[0,4]$} & $4.0$ & $4.0$ & $4.0$\tabularnewline
\cline{4-6} 
 &  &  & $0.24$ & $0.0$ & $0.0$\tabularnewline
\cline{2-6} 
 & \multirow{2}{*}{$C_{0}=0.52$} & \multirow{2}{*}{$[10^{-4},10]^{*}$ } & $1.4$ & $1.88$ & $2.22$\tabularnewline
\cline{4-6} 
 &  &  & $10^{-4}$ & $10^{-4}$ & $10^{-4}$\tabularnewline
\cline{2-6} 
 & \multirow{2}{*}{$\tau_Y=2.72$} & \multirow{2}{*}{$[0.01,100]^{*}$ } & $100$ & $100$ & $100$\tabularnewline
\cline{4-6} 
 &  &  & $1.02$ & $0.74$ & $0.62$\tabularnewline
\hline 
\multicolumn{6}{|c|}{}\tabularnewline
\hline 
\multirow{4}{*}{III.b} & \multirow{2}{*}{$r_{\nu N}=0.35$} & \multirow{2}{*}{$[0,1]$ } & $0.99$ & $1.0$ & $1.0$\tabularnewline
\cline{4-6} 
 &  &  & $0.1$ & $0.03$ & $0.01$\tabularnewline
\cline{2-6} 
 & \multirow{2}{*}{$\tau_Y=0.88$} & \multirow{2}{*}{$[0.01,100]^{*}$ } & $2.41$ & $6.65$ & $13.09$\tabularnewline
\cline{4-6} 
 &  &  & $0.48$ & $0.36$ & $0.31$\tabularnewline
\hline 
\multicolumn{6}{|c|}{}\tabularnewline
\hline 
\multirow{4}{*}{IV.b} & \multirow{2}{*}{$r_{\nu N}=0.23$} & \multirow{2}{*}{$[0,1]$ } & $0.91$ & $1.0$ & $1.0$\tabularnewline
\cline{4-6} 
 &  &  & $0.0$ & $0.0$ & $0.0$\tabularnewline
\cline{2-6} 
 & \multirow{2}{*}{$\tau_Y=1.13$} & \multirow{2}{*}{$[0.01,100]^{*}$} & $2.64$ & $5.94$ & $11.54$\tabularnewline
\cline{4-6} 
 &  &  & $0.64$ & $0.48$ & $0.41$\tabularnewline
\hline 
\multicolumn{6}{|c|}{}\tabularnewline
\hline 
\multirow{2}{*}{V} & \multirow{2}{*}{$\tau_Y=1.9$} & \multirow{2}{*}{$[0.01,100]^{*}$ } & $3.67$ & $8.71$ & $16.56$\tabularnewline
\cline{4-6} 
 &  &  & $1.13$ & $0.74$ & $0.6$\tabularnewline
\hline 
\end{tabular}
\par\end{centering}
\caption{\label{tab:Confidence-level-intervals}Confidence level
  intervals. The lifetime $\tau_Y$ is in units of $\times 10^{28}$
  s, the normalization of the power-law $C_{0}$ is in units of
  GeV cm$^{-2}$ sr$^{-1}$ s$^{-1}$ and $r_{\nu N}$ is the branching
  ratio of the channel $Y \rightarrow$ $\nu N$.  The intervals with
  format $[a,b]^{*}$ are scanned in logarithmic scale.}
\end{table}

\bibliographystyle{JHEP}
\bibliography{./icecube_nu_JHEP}

\end{document}